\documentclass[aps,prd,superscriptaddress,showpacs, preprint]{revtex4-1}
\usepackage{graphicx}
\usepackage{epstopdf}
\usepackage{amsmath}
\usepackage{amsfonts} 
\usepackage{amssymb}
\usepackage{hyperref}
\usepackage[english]{babel}
\usepackage[utf8]{inputenc}
\usepackage{xcolor}
\usepackage{bm}
\usepackage{float}

\begin{document}
\title{Quantum corrected gravitational  potential beyond monopole-monopole interactions}

\author{G.P. de Brito}\email{gpbrito@cbpf.br}
\affiliation{Centro Brasileiro de Pesquisas F\'{i}sicas (CBPF), Rua Dr. Xavier Sigaud 150, Urca, Rio de Janeiro, Brazil, CEP 22290-180}

\author{M.G. Campos}\email{mgcampos@cbpf.br}
\affiliation{Centro Brasileiro de Pesquisas F\'{i}sicas (CBPF), Rua Dr. Xavier Sigaud 150, Urca, Rio de Janeiro, Brazil, CEP 22290-180}

\author{L.P.R. Ospedal}\email{leoopr@cbpf.br}
\affiliation{Centro Brasileiro de Pesquisas F\'{i}sicas (CBPF), Rua Dr. Xavier Sigaud 150, Urca, Rio de Janeiro, Brazil, CEP 22290-180}

\author{K.P.B. Veiga}\email{kimveiga@ifba.edu.br}
\affiliation{Instituto Federal da Bahia (IFBA) - Campus Sim\~{o}es Filho, Via Universitaria s/n, Pitanguinhas, Sim\~{o}es Filho, BA, Brazil, CEP 43700-000}

\begin{abstract}

We investigate  spin- and velocity-dependent contributions to the gravitational inter-particle potential. The methodology adopted here is based on the expansion of the effective action in terms of form factors encoding quantum corrections. Restricting ourselves to corrections up to the level of the graviton propagator, we compute, in terms of general form factors, the non-relativistic gravitational potential associated with the scattering of spin-0 and -1/2 particles. We discuss comparative aspects concerning different types of scattered particles and  we also establish some comparisons with the case of electromagnetic potentials.  Moreover, we apply our results to explicit examples  of form factors based on non-perturbative approaches for quantum gravity.  Finally, the cancellation of Newtonian singularity is analysed in the presence of  terms beyond the monopole-monopole sector.

\end{abstract}



\maketitle

\section{Introduction} \label{Sec_Intro}
\indent

The current paradigm in the description of the gravitational interaction has foundation in Einstein's general relativity (GR), that describes gravity as a classical field theory for the space-time dynamics. The other known fundamental interactions are very well described in terms of quantum field theory (QFT), culminating in the standard model of particle physics. Combining gravity with the other fundamental interactions remains as one of the most challenging tasks in theoretical physics. In particular, a completely (self-)consistent theory of quantum gravity is still missing. 

Since the space-time metric plays the role of a dynamical variable in GR, a direct approach entails a QFT treatment to the quantization of metric fluctuation around a fixed background \cite{Kiefer_book,Percacci_book}. This approach, sometimes referred as covariant quantum gravity, was readily identified as a problematic QFT due to appearance of ultraviolet (UV) divergences that could not be absorbed by standard (perturbative) renormalization techniques. This problem, however, should not be taken as a dead end for the covariant quantum gravity approach. 
\begin{itemize}
    \item The most immediate way out to this problem relies on the interpretation of this approach as an effective field theory (EFT) \cite{Donoghue_EFT_QG}, which provides a consistent framework for quantum gravity calculations  valid below some cutoff scale $\Lambda_{\textmd{QG}}$.
    \item The problem of perturbatively non-renormalizable interactions can be circumvented by the inclusion of curvature squared terms in the action describing the gravitational dynamics \cite{Stelle}. This approach, however, seems to imply unitarity violation (and instabilities, at the classical level) due to the appearance of higher-derivative terms. In the last few years, the interest in theories with higher curvature terms was renewed with some interesting ideas that might conciliate unitarity and (perturbative) renormalizability within this framework (see, for example, Refs. \cite{Holdom,Modesto,ModestoShapiro,Donoghue+Menezes,Anselmi}). 
    \item Beyond the perturbative paradigm, the asymptotic safety program for quantum gravity \cite{Weinberg_ASQG,Reuter_PRD} has been investigated as a candidate for a consistent UV complete scenario for covariant quantum gravity. In this context, UV completion is achieved as a consequence of quantum scale-symmetry emerging as result of a possible fixed point in the renormalization group flow. By now, there is vast collection of results indicating the viability of this scenario \cite{Percacci_book,Reuter_Book} including possible phenomenological consequences (see the reviews \cite{Astrid_review1,Astrid_review2} and references therein).
\end{itemize}

 A consistency check in quantum gravity models based on standard QFT techniques is the investigation of quantum corrections to the Newtonian potential.  This question was originally addressed in the seminal paper by Donoghue within the EFT approach for quantum gravity \cite{Donoghue_EFT_QG}. Since then EFT and other methods have been used by several authors to carry out quantum gravitational corrections to the inter-particle potentials (see, for example, Refs. \cite{MV_PRD52,HL_PLB357,ABS_PLB395,KK_JETP95,Bjerrum_PRD66,BDH_PRD67,Faller_PRD}). Although the usual research of non-relativistic potentials concentrates in the monopole-monopole sector, a series of works in the literature also consider the contributions of spin and velocity.  In this case, spin-orbit and spin-spin interactions may appear. For instance, in Ref. \cite{gupta1966} the authors calculated the potentials related to one-graviton exchanged between particles with different spins. Long-range gravitational potentials and its spin-dependent interactions were obtained in Refs. \cite{KK_JETP98,K_NPB728,RH_JPA,HR_0802.0716} by taking into account gravitational scattering at one-loop approximation within the EFT formalism. In a similar way, the spin contributions of one-loop diagrams with mixed gravitational-electromagnetic scattering were investigated in Refs. \cite{Butt_PRD74,HR_0802.0717}. For reviews of theoretical and experimental researches on the role of spin in gravity, we point out Refs. \cite{wtNi, wtNi2}.

 In this work  we investigate spin- and velocity-dependent contribution to the gravitational inter-particle potential within a framework motivated by quantum gravity models. Our main goal is to present a detailed discussion  on the structure of possible quantum corrections to each sector beyond the monopole-monopole interaction. For this purpose, we  combine the effective action  formalism with  an expansion in terms of form factors to introduce quantum corrections at the level of the graviton propagator.  This strategy allows us to explore structural aspects of spin- and velocity-dependent contributions without relying in any specific perturbative calculation. 

This paper is organized as follows: in Section \ref{Sec_Potentials}, we present our methodology and carry out the inter-particle gravitational potentials for interactions involving spin-0 or spin-1/2 external particles in terms of general form factors. After that,  we analyse each sector  beyond monopole-monopole interaction and discuss the comparative aspects between spin-0 and spin-1/2 cases.  In addition, we also establish comparisons with the inter-particle potentials mediated by electromagnetic interaction. In Section \ref{Sec_App}, we apply our results  to particular examples motivated by non-perturbative approaches to quantum gravity.  Next, in Section \ref{Singularities}, we discuss some aspects related to the cancellation of Newtonian singularities in  higher-derivative gravity models. Finally, in Section \ref{Sec_Concluding}, we present our concluding remarks and perspectives. In the Appendix, we display some useful integrals and definitions. Throughout this work we adopted natural units where $\hbar = c = 1$, the Minkowski metric with signature $(+,-,-,-)$. The Riemann and Ricci curvature tensors were defined as $R^{\mu}_{\,\,\,\nu\alpha\beta} = \partial_\alpha \Gamma^\mu_{\nu\beta} + \Gamma^{\mu}_{\alpha\lambda} \Gamma^{\lambda}_{\nu\beta} - (\alpha\leftrightarrow\beta)$  and $R_{\mu \nu} = R^{\alpha}_{\,\,\,\mu\alpha\nu}$, respectively.

\section{Non-relativistic potentials} \label{Sec_Potentials}
\indent 

\begin{figure}[ht]
	\begin{center}
		\leavevmode
		\includegraphics[width=0.3\textwidth]{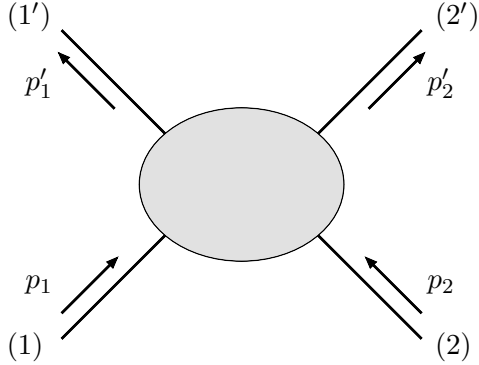}
		\put(-159,-5){(1)}
		\put(-152,20){$p_1$}
		\put(3,-5){(2)}
		\put(0,20){$p_2$}
		\put(-159,120){($1^\prime$)}
		\put(-152,95){$p_1^\prime$}
		\put(3,120){($2^\prime$)}
		\put(0,95){$p_2^\prime$}
	\end{center}
	\caption[Feynman diagram 1]{\footnotesize{Representation of a process with particles labeled by $1$ and $2$ scattering into final states labeled by $1^\prime$ and $2^\prime$. The arrows indicate the momenta assignments adopted in this paper.}}
    \label{Scattering-Process}
\end{figure}

 Let us initially introduce the methodology adopted for computing inter-particle potentials and present the approximations we are dealing with. In order to obtain spin- and velocity-dependent contributions to non-relativistic  (NR) potentials mediated by gravity, we employ the first Born-approximation, namely

\begin{equation}
V(r)= -\int \frac{d^3 \vec{q}}{(2\pi)^3} \mathcal{M}_{_{\textmd{NR}}} (\vec{q}) \, e^{i \vec{q} \cdot \vec{r}} \, ,
\label{pot_def} 
\end{equation}
where $\mathcal{M}_{_{\textmd{NR}}} (\vec{q})$ indicates the NR limit of the Feynman amplitude, $\mathcal{M}$, associated with the process $1+2 \to 1^\prime + 2^\prime$ represented in Fig. \ref{Scattering-Process}. Following Ref. \cite{livro_Maggiore}, we note that the  NR limit involves an appropriate normalization factor such that

\begin{equation}
\mathcal{M}_{_{\textmd{NR}}} (\vec{q}) = {\lim}_{_{\textmd{NR}}} \prod_{i=1,2} (2E_i)^{-1/2} \prod_{j=1,2} (2E'_j)^{-1/2} \,  \mathcal{M} (\vec{q}) \, .
\label{prescription} 
\end{equation}

The most direct way to include quantum corrections to the NR gravitational potential relies on the perturbative approach. In this case, the amplitude associated with the process in Fig. \ref{Scattering-Process} involves all the connected Feynman diagrams up to a fixed order in perturbation theory. This approach has been successfully applied to the computation of quantum corrections to the gravitational inter-particle potential in the context of EFT \cite{Donoghue_EFT_QG,HL_PLB357,ABS_PLB395,KK_JETP95,Bjerrum_PRD66,BDH_PRD67,Faller_PRD,KK_JETP98,K_NPB728,RH_JPA,HR_0802.0716}.

Alternatively, one can think in terms of the effective action formalism. In this case, the amplitude associated with the process represented in Fig. \ref{Scattering-Process} is constructed as a sum over connected ``tree-level'' diagrams with propagator and vertices extracted from the effective action $\Gamma$. The typical evaluation of the effective action $\Gamma$ relies on perturbative methods and, therefore, produce equivalent results with respect to the approach described in the previous paragraph.


The effective action formalism might be useful in order to access information beyond the perturbative approach. For example, in Ref. \cite{Knorr}, Knorr and Saueressig proposed the reconstruction of an effective action for quantum gravity starting from non-perturbative data obtained via causal dynamical triangulation. Furthermore, the effective action is expanded in terms of form factors carrying (non-)perturbative quantum corrections. For a recent discussion on form factors for quantum gravity in connection with functional renormalization group methods, see Refs. \cite{Bosma_PRL_123, Knorr_Form_Factors}. 

\begin{figure}[ht]
	\begin{center}
		\leavevmode
		\includegraphics[width=0.5\textwidth]{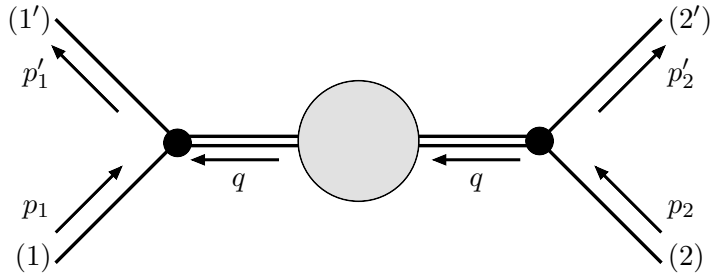}
		\put(0,0){(2)}
		\put(0,20){$p_2$}
		\put(-247,0){(1)}
		\put(-244,20){$p_1$}
		\put(0,90){($2^\prime$)}
		\put(0,70){$p_2^\prime$}
		\put(-250,90){($1^\prime$)}
		\put(-244,70){$p_1^\prime$}
		\put(-75,30){$q$}
		\put(-165,30){$q$}
	\end{center}
	\caption[Feynman diagram 3]{Diagrammatic representation of the approximation done in this paper. The arrow indicate the momentum assignments adopted in the calculation of the scattering process.}
	\label{Diagrams_3}
\end{figure}

In this paper we combine the effective action formalism with an expansion in terms of form factors in order to include quantum corrections on the NR inter-particle gravitational potential beyond monopole-monopole interactions. As a first approach we include only quantum corrections to the graviton propagator. In this case, the relevant contribution to the process depicted in Fig. \ref{Scattering-Process} corresponds to the diagram represented in Fig. \ref{Diagrams_3}.  Within this approximation, quantum corrections to the vertices are not considered  and the relativistic amplitude takes the form
\begin{align} \label{Scattering_Amp_General}
i\mathcal{M} = i\,T^{\mu\nu}(p_1,p_1^\prime) \, 
\langle h_{\mu\nu}(-q) h_{\alpha\beta}(q)\rangle \, i\,T^{\alpha\beta}(p_2,p_2^\prime) \,
\end{align}
where $T^{\mu\nu}$ stands for the tree-level energy momentum tensor associated with the scattered particles and $\langle h_{\mu\nu}(-q) h_{\alpha\beta}(q)\rangle$ denotes the graviton full-propagator.  The fact that we are not taking into account quantum corrections to the vertex imposes some limitation in the range of validation of our results. In particular, there is no \textit{a priori} reason to argue that vertex corrections should be suppressed in our investigation. In this sense, the approach adopted here should be interpreted as a first step towards the inclusion of non-perturbative effects, encoded in a form factor expansion, to the NR gravitational potential with contributions beyond the static regime. In principle, vertex corrections can also be implemented in a form factor expansion \cite{Knorr_Form_Factors,Draper}, however, this goes beyond the purposes of the present work. 

Our purpose is not to compute the effective action for quantum gravity. Instead, we assume a ``template'' for the effective action expanded in terms of form factors and motivated by symmetry arguments.
In general gauge theories, the effective action typically takes the form $\Gamma = \bar{\Gamma} + \hat{\Gamma}$, where $\delta_{\textmd{gauge}} \bar{\Gamma} = 0$ and $\delta_{\textmd{gauge}} \hat{\Gamma} \neq 0 $. Nevertheless, the ``symmetry breaking'' contribution $\hat{\Gamma}$ is controlled by Slavnov-Taylor identities for $\Gamma$. The covariant approach for quantum gravity, thought as a QFT for the fluctuation field $h_{\mu\nu}$ around a fixed background with metric $\bar{g}_{\mu\nu}$, could be faced as a gauge theory for diffeomorphism transformations. In this case, a template for the effective action in quantum gravity should take the form
\begin{align}
\Gamma[h;\bar{g}] = \bar{\Gamma}[g] + \hat{\Gamma}[h;\bar{g}] \,,
\end{align}
where $\delta_{\textmd{diff.}} \bar{\Gamma} = 0$ and $\delta_{\textmd{diff.}} \hat{\Gamma} \neq 0$. We note that the symmetric part, $\bar{\Gamma}[g]$, depends only on the full metric $g_{\mu\nu}$, while the ``symmetry breaking'' sector presents separated dependence on $\bar{g}_{\mu\nu}$ and $h_{\mu\nu}$.  In  the present paper the fluctuation field $h_{\mu\nu}$ was defined in terms of the linear split $g_{\mu\nu} = \bar{g}_{\mu\nu}+\kappa h_{\mu\nu}$ (with $\kappa = \sqrt{32\pi G}$). In this case, there is an additional local symmetry, namely split symmetry, corresponding to the combined transformation $\delta_{\textmd{split}} h_{\mu\nu}(x)=\kappa^{-1} \epsilon(x)$ and $\delta_{\textmd{split}} \bar{g}_{\mu\nu}(x)=-\epsilon(x)$ that leaves the full metric invariant $\delta_{\textmd{split}} g_{\mu\nu}= 0$ and, therefore, $\delta_{\textmd{split}}\bar{\Gamma}[g] = 0$. However, the separated dependence of $\bar{g}_{\mu\nu}$ and $h_{\mu\nu}$ in $\hat{\Gamma}[h;\bar{g}]$ implies $\delta_{\textmd{split}}\Gamma[h;\bar{g}] \neq 0$, leading to non-trivial Nielsen identities (or split Ward identities). 

For the symmetric part, we consider a template for the effective action organized in terms of a curvature expansion, given by
\begin{equation}
\bar{\Gamma}[g_{\mu \nu}] = 
\frac{2}{\kappa^2} \int d^4 x \, \sqrt{-g} \left( -2 \Lambda -  R - \frac{1}{3} R F(\Box) R + C_{\mu \nu \alpha \beta} W(\Box) C^{\mu \nu \alpha \beta} \right) + \mathcal{O}(\mathcal{R}^3) \,,
\label{eff_action} 
\end{equation}
where $\Lambda$ and $C^{\mu \nu \alpha \beta}$ denote the cosmological constant and Weyl tensor, respectively, while $F(\Box)$ and $W(\Box)$ correspond to form factors encoding quantum corrections contributing to the curvature squared sector.
Furthermore, $\mathcal{O}(\mathcal{R}^3)$ indicates all other contributions composed by curvature invariant with power higher than two. For the explicit computations performed in this paper, we consider flat background metric, i.e., $g_{\mu \nu} = \eta_{\mu \nu} + \kappa \, h_{\mu \nu}$. In this case, the relevant contributions for the full graviton propagator come exclusively from terms up to $\mathcal{O}(\mathcal{R}^2)$.

For the symmetry breaking sector, we use a template with the same functional form as a typical gauge fixing term added to classical action, namely
\begin{align}
\hat{\Gamma}[h_{\mu \nu};\bar{g}] = \frac{1}{2\alpha} \int d^4x 
\sqrt{-\bar{g}} \,\bar{g}^{\mu\nu} F_\mu [h;\bar{g}] F_\nu[h;\bar{g}] \,,
\end{align}
where $F_\mu[h;\bar{g}] = \bar{\nabla}^\nu h_{\mu\nu} - \frac{1}{2} \bar{\nabla}_\mu h  $.  
 One can argue that a different choice in this sector would not affect our results since we are computing gauge-independent quantities (on-shell amplitudes) and, therefore, any gauge-dependence should drop out in the final results.

Bearing in mind our template for the effective action, the graviton ``full''-propagator (around flat background) is readily computed as the inverse of the 2-point function $\delta^2 \Gamma/\delta h^2|_{h=0}$, resulting in the following expression
\begin{eqnarray}
\langle h_{\mu \nu} (-q) h_{\alpha \beta} (q) \rangle = 
\frac{i}{q^2} \Bigg[ \frac{1}{Q_2 (q^2)} \mathcal{P}_{\mu\nu\alpha\beta}^{(2)}  - \frac{1}{2Q_0 (q^2)} \mathcal{P}_{\mu\nu\alpha\beta}^{(0)}   \Bigg] \, + \,
i\Delta_{\mu \nu \alpha \beta}(q) \,,
\label{prop_simpl} 
\end{eqnarray}
where we define 
\begin{subequations}
\begin{align}\label{def_Q2}
Q_2 (q^2 ) = 1 +  \frac{2 \Lambda}{q^2} + 2 q^2 \, W(-q^2) \, ,
\end{align}
\begin{align}\label{def_Q0}
Q_0 (q^2 ) = 1 +  \frac{2 \Lambda}{q^2} + 2 q^2 \, F(-q^2) \, .
\end{align}
\end{subequations}
In addition, the tensor structures $\mathcal{P}_{\mu\nu\alpha\beta}^{(2)} $ and $\mathcal{P}_{\mu\nu\alpha\beta}^{(0)} $ are defined as
\begin{subequations}
\begin{align}
\mathcal{P}_{\mu\nu\alpha\beta}^{(2)} =
\frac{1}{2} (\eta_{\mu \alpha} \eta_{\nu \beta}+ \eta_{\mu \beta} \eta_{\nu \alpha}) - \frac{1}{3}\eta_{\mu \nu}\eta_{\alpha \beta} \,,
\end{align}
\begin{align}
\mathcal{P}_{\mu\nu\alpha\beta}^{(0)} = \frac{1}{3}\eta_{\mu \nu}\eta_{\alpha \beta} \,.
\end{align}
\end{subequations}
The remaining terms in the graviton propagator, represented by $i\Delta_{\mu \nu \alpha \beta}(q)$,  vanish when contracted with the energy-momentum tensor of the scattered particles.

It is worthwhile mentioning that using the effective action \eqref{eff_action}, where form factors $F(\Box)$ and $W(\Box)$ are introduced 
with scalar curvature and Weyl tensor, we obtain the propagator \eqref{prop_simpl} in which the contributions of these form factors are disconnected. In other words, from Eqs. \eqref{def_Q2} and \eqref{def_Q0}, we observe that $F(\Box)$ and $W(\Box)$ contribute only to scalar and graviton modes, respectively.  

In what follows we present our results for the NR gravitational potential, taking into account the scattering of both massive spin-0 and spin-1/2 particles, with quantum corrections being included in terms of general form factors $F(\Box)$ and $W(\Box)$. As usually done in the literature of spin- and velocity-dependent potentials, we adopt the center-of-mass (CM) reference frame, described in terms of the $3-$momentum transfer $\vec{q}$ and average momentum $\vec{p}$. The CM variables are related to the momentum assignments depicted in Fig. \ref{Diagrams_3} in terms of the following expressions
\begin{equation}
\vec{p}_1 = -\vec{p}_2 = \vec{p} - \frac{\vec{q}}{2} \, , \qquad   
\vec{p'}_1 = -\vec{p'}_2 = \vec{p} + \frac{\vec{q}}{2} \,  .
\label{momentum_CM} \end{equation}




Since we are dealing with an elastic scattering, the total energy of the system is conserved. With this assumption and using the momentum attributions in \eqref{momentum_CM}, it is possible to show that $ \vec{q} \cdot \vec{p} = 0$. This result implies $ E_1 = E'_1 $ and $E_2 = E'_2 $ or equivalently $ q^\mu = (0,\vec{q})$. In the non-relativistic limit we take $m_i^2 >> \vec{q}^{\, 2}, \vec{p}^{\, 2}$, leading to the approximation  $E_i \approx m_i + \frac{1}{2m_i} (\vec{p}^{\,2} + \vec{q}^{\,2}/4)$. In what follows we apply these conditions and approximations to the energy-momentum tensor appearing in Eq. \eqref{Scattering_Amp_General} to arrive at the non-relativistic amplitude. This prescription defined directly in terms of CM variables is equivalent to an expansion in powers of $\vec{p}_i/m_i$ and $\vec{p}_i^{\,\,\prime}/m_i$. In our calculations we consider contributions up to second order in these expansion variables.

\subsection{Spin-0 external particles} 
\label{spin_0_subsection}
\indent 



Within the working setup above described, we first investigate the case of gravitationally interacting spin-0 particles.  The investigation performed in this paper takes into account an approach where quantum correction to the vertices are neglected. In this sense, our template for the effective action in the spin-0 sector essentially corresponds to the classical action of a scalar field minimally coupled to gravity, 
\begin{eqnarray}
\Gamma_{\textmd{scalar}}[\phi,g] = \int{d^4x \, \sqrt{-g} \Bigg( \frac{1}{2} g^{\mu\nu} \partial_\mu \phi \partial_\nu \phi - \frac{1}{2}m^2 \phi^2} \Bigg) \, .
\end{eqnarray} 
 By expanding up to first order in the fluctuation field $h_{\mu\nu}$, we can directly extract the Fourier representation of the energy momentum tensor associated with the external legs in Fig. \ref{Diagrams_3}, namely
\begin{align}\label{scalar_vertex}
T_{\mu \nu}(p, p') = -\frac{\kappa}{2} \Big(\, p_\mu p'_\nu + p_\nu p'_\mu - \eta_{\mu \nu} \left(  p \cdot p' - m^2 \right) \Big) \, .
\end{align}
We adopt conventions where the momenta $p$ and $p'$ are respectively assigned as incoming and outgoing with respect to the vertex.

The relativistic scattering amplitude is computed in terms of Eq. \eqref{Scattering_Amp_General} along with Eqs. \eqref{prop_simpl} and \eqref{scalar_vertex}.  We note that the external legs in Fig. \ref{Diagrams_3} are considered to be on-shell and, therefore, $p_1^2 = p_1^{\prime 2} = m_1^2$ and $p_2^2 = p_2^{\prime 2} = m_2^2$. In fact, using the conservation of the energy-momentum tensor ($q_\mu \, T^{\mu\nu}(p_i,p_i^\prime) = 0$, for on-shell in- and out-states) we arrive at the intermediary result
\begin{eqnarray}\label{amp_esc_1} 
i \mathcal{M}^{(s=0)} =  \frac{i}{q^2} \left[ \left( 
\frac{1}{3 Q_2} + \frac{1}{6 Q_0} \right) T_{1 \, \, \mu}^\mu T_{2 \, \, \beta}^\beta  
-  \frac{1}{Q_2} \, T_1^{\mu \nu} T_{2 \, \mu \nu} \right] \, ,
\end{eqnarray}
where we work with the shorthand notations $ T^{\mu \nu}_i \equiv T^{\mu \nu}(p_i, p'_i) $ and $Q_i \equiv Q_i (q^2)$. After some simple algebraic manipulations using the explicit expression for the energy-momentum tensor we find the following result for the scattering amplitude 

\begin{eqnarray}\label{amp_esc_4} 
 \mathcal{M}^{(s=0)} &=& \frac{\kappa^2}{6 q^2 \, Q_2} \Bigg( 2 m_1^2 m_2^2 - 3 (p_1 \cdot p_2) (p_1' \cdot p_2') - 3 (p_1 \cdot p_2') (p_1' \cdot p_2)  \nonumber \\
&+&  2 (p_1 \cdot p_1') (p_2 \cdot p_2') - m_1^2 \, p_2 \cdot p_2' - m_2^2 \, p_1 \cdot p_1' \Bigg)  \nonumber \\
&+& \frac{\kappa^2}{6 q^2 \, Q_0} \Bigg( (p_1 \cdot p_1') (p_2 \cdot p_2') - 2 m_1^2 \, p_2 \cdot p_2' 
- 2 m_2^2 \, p_1 \cdot p_1' + 4 m_1^2 m_2^2 \Bigg) \, .
\end{eqnarray}

In order to obtain the NR description, we use the prescription \eqref{prescription}. In the CM reference frame with momentum attributions \eqref{momentum_CM}, we have
\begin{eqnarray}
\mathcal{M}^{(s=0)}_{\textrm{NR}} &=& \frac{ \kappa^2 m_1 m_2 }{6 \, Q_2 \, \vec{q}^{\,2}} \, \left\{ 1 +  \vec{p}^{\,2} \left( \frac{3}{m_1 m_2} + \frac{1}{m_1^2} + \frac{1}{m_2^2} \right)
+ \frac{ \vec{q}^{\,2} }{8} \,  \left( \frac{1}{m_1^2} + \frac{1}{m_2^2}  \right) + \mathcal{O}(3) \right\}  \nonumber \\
&-& \frac{ \kappa^2 m_1 m_2 }{24 \, Q_0 \, \vec{q}^{\, 2} } \, \left\{ 1  - \frac{\vec{p}^{\, 2} }{2} \left( \frac{1}{m_1^2} + \frac{1}{m_2^2} \right) - \frac{5 \, \vec{q}^{\, 2}}{8}  \left( \frac{1}{m_1^2} + \frac{1}{m_2^2} \right) + \mathcal{O}(3)\right\}
\, ,\label{amp_NR_esc} \end{eqnarray}
 with $\mathcal{O}(3)$ indicating terms higher than second order in $|\vec{p}|/m_{1,2}$ and/or $|\vec{q}|/m_{1,2}$, which we shall neglect.

Finally, by taking the Fourier integral, Eq. \eqref{pot_def}, we promptly obtain the  inter-particle gravitational potential  with contributions beyond the monopole-monopole sector
\begin{eqnarray}
V^{(s=0)}(r) &=& - \frac{\kappa^2 m_1 m_2}{6}  \Bigg\{ I_1^{(2)}(r) + \vec{p}^{\,2} \left( \frac{3}{m_1 m_2} + \frac{1}{m_1^2} + \frac{1}{m_2^2} \right) I_1^{(2)}(r) \nonumber \\
&+&  \frac{1}{8} \left( \frac{1}{m_1^2} + \frac{1}{m_2^2} \right) I_0^{(2)}(r) \Bigg\} 
+  \frac{\kappa^2 m_1 m_2}{24}  \Bigg\{  I_1^{(0)}(r) \nonumber \\
&-& \frac{ \vec{p}^{\,2} }{2} \left(  \frac{1}{m_1^2} + \frac{1}{m_2^2} \right)  I_1^{(0)}(r) - \frac{5}{8} \left( \frac{1}{m_1^2} + \frac{1}{m_2^2} \right) I_0^{(0)}(r) \Bigg\}   , 
\label{spin_0_pot}
\end{eqnarray}
where the integrals $I_n^{(a)}(r)$ are defined in Appendix, Eq. \eqref{I_a_n} with $n=0,1$ and $a=0,2$.

\subsection{Spin-1/2 external particles} 
\label{spin_meio_subsection}
\indent 




In present subsection, we describe the  gravitational interaction between two spin-1/2 particles.  Since we do not take into account any vertex correction, our template for this sector basically corresponds to the classical action of the Dirac field minimally coupled to gravity, namely
\begin{eqnarray}
\Gamma_{\textmd{ferm}}[\bar{\psi},\psi,g] = \int{d^4x \, \sqrt{-g} \left( \frac{i}{2} (\bar{\psi} \, \gamma^\mu_g \, \nabla_\mu \psi - \nabla_\mu \bar{\psi} \, \gamma^\mu_g  \, \psi ) - m\bar{\psi} \psi \right)} \, .
\end{eqnarray}
To define the fermion in a curved space-time we use the spin-base formalism, where the covariant derivative is defined according to $\nabla_\mu  \psi = \partial_\mu \psi + \Gamma_\mu \psi$ and $\nabla_\mu \bar{\psi} = \partial_\mu \bar{\psi} - \bar{\psi}\,\Gamma_\mu$, with $\bar{\psi}$ and $\Gamma_\mu$ representing the Dirac conjugate and an appropriate connection, respectively (see Refs. \cite{spinbase1,spinbase2,spinbase3} for more details). In addition, the matrices $\gamma^\mu_g$ satisfy the Clifford algebra $\left\{ \gamma^\mu_g , \gamma^\nu_g \right\} = 2 g^{\mu \nu} \bm{1}$. The tree-level vertex involving two fermions and one graviton is extracted by expanding $\Gamma_{\textmd{ferm}}[\bar{\psi},\psi,g]$ up to first order in the fluctuation field $h_{\mu\nu}$. Based on the resulting expression, we can obtain the energy-momentum tensor 
\begin{align}\label{fermion_vertex}
T_{\mu \nu}(p, p') =& \,
\frac{\kappa}{8} \Big( 2 \eta_{\mu \nu} \big( (p +p')_\alpha \,\mathcal{J}^\alpha(p, p') - 2m \, \rho(p, p')\big)  \nonumber \\
&\,\,-  (p + p')_\mu \mathcal{J}_\nu(p, p') - (p + p')_\nu \mathcal{J}_\mu(p, p') \Big) \, .
\end{align}
We define the bi-linear structures $\mathcal{J}^\mu(p, p') = \bar{u}(p') \gamma^\mu u(p)$ and $\rho(p, p') = \bar{u}(p') u(p)$, where $u(p)$ denotes the free positive  energy solution for the four-component spinor and $\bar{u}(p)={u}^\dagger(p) \gamma^0$. Here, it should be noted that $\gamma^\mu$ corresponds to the usual gamma matrices  in a flat background, satisfying $\left\{ \gamma^\mu, \gamma^\nu \right\} = 2 \eta^{\mu \nu} \bm{1}$. Then, combining Eqs. \eqref{Scattering_Amp_General} and \eqref{prop_simpl}, we arrive in a similar expression as in the case of spin-0 particles, namely
\begin{eqnarray}\label{amp_fermion}
i \mathcal{M}^{(s=1/2)} = \frac{i}{q^2} \left[ \left( \frac{1}{3 Q_2} + \frac{1}{6 Q_0} \right) T_{1 \, \mu}^{\,\, \mu} T_{2 \, \beta}^{\,\, \beta} -   \frac{1}{Q_2} T_{1}^{\, \mu \nu} T_{2 \, \mu \nu} \right] \, .
\end{eqnarray}

Expanding the energy-momentum tensor in terms of the  bi-linears $\mathcal{J}^\mu$ and $\rho$, we find the relativistic scattering amplitude

\begin{eqnarray}
\mathcal{M}^{(s=1/2)} &=& \frac{ \kappa^2}{q^2 Q_2} \Bigg\{ \frac{1}{16} (p_1 + p_1')_\mu (p_2 + p_2')_\nu \mathcal{J}_1^{\mu}  \mathcal{J}_2^{\nu}  - \frac{m_1}{8} \rho_1  (p_2 + p_2')_\mu \mathcal{J}_2^{\mu} - \frac{m_2}{8} \rho_2  (p_1 + p_1')_\mu \mathcal{J}_1^{\mu}\nonumber \\
&-& \frac{1}{32} (p_1 + p'_1)^\nu (p_2 + p'_2)_\nu \mathcal{J}_1^{\mu}  
\mathcal{J}_{2\mu}- \frac{1}{32}   (p_1 + p_1')_\mu (p_2 + p_2')_\nu \mathcal{J}_2^{\mu}  \mathcal{J}_1^{\nu}  + \frac{m_1 m_2}{3}  \rho_1 \rho_2 \Bigg\}    \nonumber \\
&+& \frac{ \kappa^2}{q^2 Q_0} \, \Bigg\{ \frac{3}{32}  (p_1 + p_1')_\mu (p_2 + p_2')_\nu \mathcal{J}_1^{\mu}  \mathcal{J}_2^{\nu}  + \frac{2 m_1 m_2}{3}  \rho_1 \rho_2  \nonumber \\
&-&  \frac{m_1}{4} \rho_1  (p_2 + p_2')_\mu \mathcal{J}_2^{\mu}  - \frac{m_2}{4}  \rho_2(p_1 + p_1')_\mu \mathcal{J}_1^{\mu}
\Bigg\} \, , 
\label{amp_fermion_rel} \end{eqnarray}
where we use the shorthand notation $\rho_j = \rho (p_j, p_j')$ and $\mathcal{J}_j^{\mu} =\mathcal{J}^{\mu} (p_j, p_j')$.


In order to extract the NR scattering amplitude,  we first remember that $u(p)$ satisfies the on-shell condition $\left[ \gamma^\mu p_\mu - m \bm{1} \right] \, u(p) = 0$. In the standard Dirac representation, we obtain
\begin{align}\label{ferm_field}
u(p) = \sqrt{E + m} \left( \begin{array}{c}\xi \\ \frac{\vec{\sigma} \cdot \vec{p}}{E + m} \, \xi\end{array}  \right) .
\end{align}
with $\xi$ and $\vec{\sigma}$ being the basic spinor and Pauli matrices, respectively. 
In the NR limit, the relevant bi-linear structures $\rho$ and $\mathcal{J}^\mu$ are written as (in the CM frame)


\begin{subequations}
    \begin{align}
    \rho_{j}|_{\textmd{NR}}  = 2\,m_j \bigg[ 1 + \frac{1}{8m_j^2} \bigg( \vec{q}^{\,2} - 4i(\vec{q} \times \vec{p}\,) \cdot \vec{S}_j \bigg) + \mathcal{O}(3) \bigg] \,,
    \end{align}
    \begin{align}
    \mathcal{J}^0_{j}|_{\textmd{NR}}  = 2\,m_j \bigg[ 1 + \frac{1}{2m_j^2} \bigg( \vec{p}^{\,2} + i(\vec{q} \times \vec{p}\,) \cdot \vec{S}_j \bigg) + \mathcal{O}(3) \bigg]  \,,
    \end{align}
    \begin{align}
    \vec{\mathcal{J}}_{j}|_{\textmd{NR}} = 2\, \chi_j \bigg[ \vec{p} - i (\vec{q} \times  \vec{S}_j) \bigg] \label{noapp} \,,
    \end{align} 
\end{subequations}
 where $j$ indicates the particle label and we have defined $\chi_1 = 1$, $\chi_2 = -1$ and the spin $\vec{S}_j = \frac{1}{2} \, \xi'^\dagger_j \vec{\sigma}\xi_j$. In addition, factors of $\xi'^\dagger_j \xi_j$ have been omitted.  As in the scalar case, the terms in $\mathcal{O}(3)$ are neglected. We highlight that there is no further approximation in Eq. (\ref{noapp}).

After some algebraic manipulations, we find that

\begin{eqnarray}
\mathcal{M}^{(s=1/2)}_{\textrm{NR}} &=& \frac{ \kappa^2 m_1 m_2 }{6 Q_2 \, \vec{q}^{\,2}} \, \Bigg\{ 1  +  \vec{p}^{\,2} \left( \frac{3}{m_1 m_2} + \frac{1}{m_1^2} + \frac{1}{m_2^2} \right)  \nonumber \\
&+& i \left[ \left( \frac{1}{m_1^2} + \frac{3}{2} \frac{1}{m_1 m_2} \right) \vec{S}_1  
+ \left( \frac{1}{m_2^2} + \frac{3}{2} \frac{1}{m_1 m_2} \right) \vec{S}_2
\right] \cdot \left( \vec{q} \times \vec{p} \, \right) \nonumber \\
&-&  \frac{3}{4} \frac{\vec{q}^{\, 2}}{m_1 m_2}  \vec{S}_1 \cdot \vec{S}_2
+ \frac{3}{4} \frac{1}{m_1 m_2} \left( \vec{q} \cdot \vec{S_1} \right) \left( \vec{q} \cdot \vec{S_2} \right) + \mathcal{O}(3) \Bigg\}
 \nonumber \\
&-& \frac{ \kappa^2 m_1 m_2 }{24 Q_0 \, \vec{q}^{\,2} } \, \Bigg\{ 1 - \frac{\vec{p}^{\,2}}{2} \left( \frac{1}{m_1^2} + \frac{1}{m_2^2} \right) \nonumber\\
&-&  \frac{i}{2} \left[ \frac{1}{m_1^2} \vec{S}_1 + \frac{1}{m_2^2} \vec{S}_2
\right] \cdot \left( \vec{q} \times \vec{p} \, \right) + \mathcal{O}(3) \Bigg\} \, .
\label{amp_NR_ferm} \end{eqnarray}

The NR gravitational potential associated with the scattering of spin-1/2 particles is obtained by performing the Fourier integral \eqref{pot_def}, resulting in the following expression

\begin{align}
&V^{(s=1/2)}(r) =  -\frac{\kappa^2 m_1 m_2}{6} \, \Bigg\{ I^{(2)}_1(r) + \vec{p}^{\,2} \left( \frac{3}{m_1 m_2} + \frac{1}{m_1^2} + \frac{1}{m_2^2} \right) I^{(2)}_1(r)    \nonumber\\
&\quad+ \left[  \left( \frac{1}{m_1^2} +\frac{3}{2} \frac{1}{m_1 m_2} \right) \vec{S}_1 +
\left( \frac{1}{m_2^2} +\frac{3}{2} \frac{1}{m_1 m_2} \right) \vec{S}_2
\right] \cdot \frac{\vec{L}}{r} \frac{d}{dr} I^{(2)}_1(r) \nonumber \\
&\quad- \frac{3}{4} \frac{\vec{S}_1 \cdot \vec{S}_2}{m_1 m_2}  \, I^{(2)}_0(r) +
 \frac{3}{4} \sum_{i,j=1}^{3} \frac{(\vec{S}_1)_i \, (\vec{S}_2)_j}{m_1 m_2} \, I_{ij}^{(2)}(r) \Bigg\} \nonumber \\
&\quad+  \frac{\kappa^2 m_1 m_2}{24} \, \Bigg\{ I^{(0)}_1(r) - \frac{\vec{p}^{\,2}}{2} \left(  \frac{1}{m_1^2} + \frac{1}{m_2^2} \right) I_1^{(0)}(r)  
-  \frac{1}{2} \left[ \frac{\vec{S}_1}{m_1^2} + \frac{\vec{S}_2}{m_2^2} \right] \cdot \frac{\vec{L}}{r} \frac{d}{dr} I^{(0)}_1(r) \Bigg\} \, ,
\label{spin_meio_pot}
\end{align}
where $\vec{L} = \vec{r} \times \vec{p}$ stands for the orbital angular momentum and the anisotropic integral $I_{ij}^{(2)}(r)$ is defined in the Appendix, Eq. \eqref{I_ij}.  The appearance of a derivative in spin-orbit interactions is related to some manipulations of the Fourier integral and spherical symmetry. For more details, see Eq. \eqref{int_A}.


\subsection{Comparative aspects of spin- and velocity-dependent potentials} 
\label{partial_conclusions_subsection}

At this stage, it is relevant to compare structural aspects of the potentials for spin-0 and spin-1/2 cases. First of all, we note that the potential for spin-0 particles is characterized by two different sectors, monopole-monopole and velocity-velocity contributions, namely
\begin{subequations}
	\begin{align}
	V_{\textmd{mon-mon}}^{(s=0)}(r) =  
	&-\frac{\kappa^2 m_1 m_2}{6} \bigg( I^{(2)}_1(r) - \frac{1}{4}  I^{(0)}_1(r) \bigg)   \, \nonumber\\
	&-\frac{\kappa^2 m_1 m_2}{48} \left( \frac{1}{m_1^2} + \frac{1}{m_2^2} \right) \bigg[I^{(2)}_0(r) 
	+\frac{5}{4}  I^{(0)}_0(r)\bigg] \,,
	\label{pot_monopole_s=0} 
	\end{align}
	\begin{align}
	V_{\textrm{vel-vel}}^{(s=0)}(r) =  -\frac{\kappa^2 m_1 m_2}{6}  \, \vec{p}^{\, 2} \Bigg\{ 
	\left( \frac{1}{m_1^2} + \frac{1}{m_2^2} \right) \left[ I^{(2)}_1(r) + \frac{1}{8} I^{(0)}_1(r) \right]
	+ \frac{3}{m_1 m_2} I^{(2)}_1(r) \Bigg\} \, .
	\label{pot_vel_s=0} 
	\end{align}
\end{subequations}
The potential associated with spin-1/2 particles, on the other hand, receives contributions from four different sectors: monopole-monopole, velocity-velocity, spin-orbit and spin-spin interactions.  These contributions are given by

\begin{subequations}
	\begin{align}
	V_{\textmd{mon-mon}}^{(s=1/2)}(r) =  
	-\frac{\kappa^2 m_1 m_2}{6} \bigg( I^{(2)}_1(r) - \frac{1}{4}  I^{(0)}_1(r) \bigg)  \,,
	\label{pot_monopole_s=1/2} 
	\end{align}
	\begin{align}
	V_{\textrm{vel-vel}}^{(s=1/2)}(r) =  -\frac{\kappa^2 m_1 m_2}{6}  \, \vec{p}^{\, 2} \Bigg\{ 
	\left( \frac{1}{m_1^2} + \frac{1}{m_2^2} \right) \left[ I^{(2)}_1(r) + \frac{1}{8} I^{(0)}_1(r) \right]
	+ \frac{3}{m_1 m_2} I^{(2)}_1(r) \Bigg\} \, ,
	\label{pot_vel_s=1/2} 
	\end{align}
	\begin{align}
	V_{\textrm{spin-orbit}}^{(s=1/2)}(r) = &-\frac{\kappa^2 m_1 m_2}{6}  \,
	\left[  \left( \frac{1}{m_1^2} +\frac{3}{2} \frac{1}{m_1 m_2} \right) \vec{S}_1 +
	\left( \frac{1}{m_2^2} +\frac{3}{2} \frac{1}{m_1 m_2} \right) \vec{S}_2
	\right] \cdot \frac{\vec{L}}{r} \frac{d}{dr} I^{(2)}_1(r) \nonumber \\
	& -\frac{\kappa^2 m_1 m_2}{48} \bigg(\frac{1}{m_1^2} \vec{S}_1+\frac{1}{m_2^2} \vec{S}_2\bigg)
	\cdot \frac{\vec{L}}{r} \frac{d}{dr} I^{(0)}_1(r) \,,
	\end{align}
	\begin{align}
	V_{\textrm{spin-spin}}^{(s=1/2)}(r) = -\frac{\kappa^2 m_1 m_2}{6}
	\bigg[ \!-\! \frac{3}{4} \frac{\vec{S}_1 \cdot \vec{S}_2}{m_1 m_2}  \, I^{(2)}_0(r) +
	\frac{3}{4} \sum_{i,j=1}^{3} \frac{(\vec{S}_1)_i \, (\vec{S}_2)_j}{m_1 m_2} \, I_{ij}^{(2)}(r) \bigg] \,.
	\end{align}
\end{subequations}

We first note that the static limit is obtained by taking the combined limit
\begin{align}
\frac{1}{m_1 m_2} V_{\textmd{stat.}}^{(s)}(r) =  
\lim_{ \substack{\vec{p}\to 0\\m_i\to \infty} } \,\frac{1}{m_1 m_2} V^{(s)}(r) .
\end{align}
In this case, the only remaining contribution comes from the monopole-monopole sector, which results in 
\begin{align}
V_{\textmd{stat.}}^{(s)}(r) =  
-\frac{\kappa^2 m_1 m_2}{6} \bigg( I^{(2)}_1(r) - \frac{1}{4}  I^{(0)}_1(r) \bigg)  \,,
\label{pot_static} 
\end{align}
both for spin-0 and spin-1/2 particles. In the particular case of vanishing form factors, i.e. without deviations from the classical Einstein-Hilbert action, we recover the usual Newtonian potential $V_{\textmd{stat.}}^{(s)}(r) = -\frac{\kappa^2 m_1 m_2}{32\pi r} \equiv -\frac{G m_1 m_2}{r}$.

Moving away from the static regime we note the similarities and differences between spin-0 and spin-1/2 cases. The monopole-monopole sectors, Eqs. \eqref{pot_monopole_s=0} and \eqref{pot_monopole_s=1/2}, contain universal contributions appearing both in the spin-0 and spin-1/2. However, as we can observe from Eq. \eqref{pot_monopole_s=0}, $V_{\textmd{mon-mon}}^{(s=0)}(r)$ has an additional term which is not present in $V_{\textmd{mon-mon}}^{(s=1/2)}(r)$. This additional term has a sub-leading behavior as we are going to see in the next section from explicit examples.

Beyond the monopole-monopole terms, we observe that the velocity-dependent sector $V_{\textrm{vel-vel}}^{(s)}(r)$ has the same form both for spin-0 and spin-1/2 cases. On the other hand, spin-orbit and spin-spin interactions are present only in the potential associated with spin-1/2 particles. While spin-orbit terms ($\sim \vec{L}\cdot \vec{S}_i$) interact via spin-2 and spin-0 graviton modes, spin-spin contributions ($\sim (\vec{S}_1)_i \, (\vec{S}_2)_j I_{ij}^{(2)} $ and $\sim \vec{S}_1 \cdot \vec{S}_2$) exhibit only interactions via spin-2 graviton modes.

It is worthy to  highlight that our methodology is applicable to modified (classical) theories of gravity with higher-order derivatives and other non-local functions. Once we have developed the potentials with arbitrary form factors, we just need to reinterpret the effective action as a classical one and redefine the $Q_0$ and $Q_2$ factors. We shall return to this point in Section \ref{Singularities}. Furthermore, we comment that, for the gravitational interaction of spin-0 particles, it is possible to generalize our results to arbitrary dimensions, as already discussed in the literature for modified theories of gravity in monopole-monopole sector (see \cite{Accioly_el_al_CQG_2015,Accioly_et_al_PRD2018} and references therein). However, for spin-$\frac{1}{2}$ case and its spin-dependent contributions, this extension shall be a non-trivial task, since the definition of spin is particular to the dimension we are dealing with. For instance, when considering space-time with odd dimension and parity symmetry (typically to electromagnetic and gravitational interactions),  a reducible representation is adoptable in order to conciliate the parity symmetry with massive fermions. In these cases, new spin-dependent effects have been discussed \cite{Dorey_Mavromatos_NPB,Leo_Helayel_PRD}. In other words, the inclusion of the spin-dependent interactions should be carefully done for each particular dimension, especially when discrete symmetries are desired. 



\subsection{Comparisons with NR electromagnetic  potentials}
\label{partial_conclusions_subsection_2}

Before we proceed with specific form factors  motivated by quantum gravity models, it is interesting to compare our results with the case of NR potentials  mediated by electromagnetic interaction. Adopting the same strategy as in the gravitational case, we consider the following template for the electromagnetic effective action
\begin{align}
\Gamma_{\textmd{EM}}[A] = -\frac{1}{4}\int d^4x \, F_{\mu\nu}(1+H(\Box))F^{\mu\nu} -
\frac{1}{2\alpha}\int d^4x \,(\partial_\mu A^\mu)^2 + \mathcal{O}(F^3) ,
\end{align}
where $H(\Box)$ denotes a form factor modeling quantum corrections up to $\mathcal{O}(A^2)$. In this case, the photon ``full''-propagator is subjected to the parameterized form 
\begin{align}
\langle A_\mu(-q) A_\nu(q) \rangle = -\frac{i}{q^2(1+H(-q^2))} \eta_{\mu\nu} + i\Delta_{\mu\nu}(q)\,,
\end{align}
where $\Delta_{\mu\nu}(q)$ indicates those contributions that vanishes when contracted with external vector currents. The photon propagator is mapped in terms of quantities defined in Ref. \cite{Gustavo_Pedro_Leo_PRD}. Therefore, we can readily import the results from \cite{Gustavo_Pedro_Leo_PRD}, leading to the following expressions
\begin{subequations}
	\begin{align}\label{EM_pot_mon-mon_s=0}
	V_{\textmd{EM, mon-mon}}^{(s=0)}(r) = e_1 e_2  I^{\textmd{EM}}_1(r) \,,
	\end{align}
	\begin{align}
	V_{\textmd{EM, vel-vel}}^{(s=0)}(r) = 
	\frac{e_1 e_2 }{m_1 m_2} \, \vec{p}^{\,2} I^{\textmd{EM}}_1(r) \,,
	\end{align}
\end{subequations}
for spin-0 particles, and 
\begin{subequations}
	\begin{align}\label{EM_pot_mon-mon_s=1/2}
	V_{\textmd{EM, mon-mon}}^{(s=1/2)}(r) =
	e_1 e_2  \left[ I^{\textmd{EM}}_1(r) - 
	\frac{1}{8} \left( \frac{1}{m_1^2} + \frac{1}{m_2^2} \right) I^{\textmd{EM}}_0(r) \right]\,,
	\end{align}
	\begin{align}
	V_{\textmd{EM, vel-vel}}^{(s=1/2)}(r) =
	\frac{e_1 e_2 }{m_1 m_2} \, \vec{p}^{\,2} I^{\textmd{EM}}_1(r)\,,
	\end{align}
	\begin{align}
	V_{\textmd{EM, spin-orbit}}^{(s=1/2)}(r) = e_1 e_2  \,
	\left[  \left( \frac{1}{2m_1^2}+\frac{1}{m_1 m_2} \right) \vec{S}_1 +
	\left( \frac{1}{2m_2^2} +\frac{1}{m_1 m_2} \right) \vec{S}_2
	\right] \cdot \frac{\vec{L}}{r} \frac{d}{dr} I_1^{\textmd{EM}}(r) ,
	\end{align}
	\begin{align}
	V_{\textrm{EM, spin-spin}}^{(s=1/2)}(r) = e_1 e_2
	\bigg[ \!-\! \frac{\vec{S}_1 \cdot \vec{S}_2}{m_1 m_2}  \, I_0^{\textmd{EM}}(r) +
	\sum_{i,j=1}^{3} \frac{(\vec{S}_1)_i \, (\vec{S}_2)_j}{m_1 m_2} \, I_{ij}^{\textmd{EM}}(r) \bigg] \,,
	\end{align}
\end{subequations}
in the case of spin-1/2 particles. The integrals $I^{\textmd{EM}}_n(r)$ and $I_{ij}^{\textmd{EM}}(r)$ follow the same definition as Eqs. \eqref{I_a_n} and \eqref{I_ij}, but replacing $Q_a(\vec{q}^{\,2})$ by $1+H(\vec{q}^{\,2})$.

As we can observe, the NR potentials mediated by electromagnetic interaction present some similarities in comparison with the gravitational case. In the monopole-monopole sector, Eqs. \eqref{EM_pot_mon-mon_s=0} and \eqref{EM_pot_mon-mon_s=1/2}, we note the appearance of universal leading order contributions (terms involving $I^{\textmd{EM}}_1(r)$) both in the case of spin-0 and spin-1/2 scattered particles. On the other hand, in contrast with the gravitational case, the additional non-universal contribution (involving $I^{\textmd{EM}}_0(r)$) appears only in the spin-1/2 case.

Beyond the monopole-monopole contribution, we first note that in the velocity-velocity sector, as in the gravitation case, exhibits the same result both for spin-0 and spin-1/2 particles. For spin-orbit and spin-spin contributions, only present in the case of spin-1/2 particles, we observe the same kind of interaction structures (terms with $\vec{S}_i\cdot \vec{L}$, $\vec{S}_1 \cdot \vec{S}_2$ and $(\vec{S}_1)_i \, (\vec{S}_2)_j I_{ij}^{(2)} $) both for electromagnetic and gravitational potentials.


\section{Form Factors Motivated by Quantum Gravity Models} \label{Sec_App}
\indent 

The results presented in the previous section carry some model independent features at the level of the graviton propagator. It allows to study structural aspects of quantum contributions to the gravitational potential beyond the monopole-monopole sector. However, a more detailed analysis depends on the evaluation of basic integrals defined in Appendix \ref{Appendix_int} for specific form factors. In what follows, we work out some examples with form factors motivated by recent investigations in the context of non-perturbative approaches for quantum gravity.

\subsection{Form factors motivated by CDT data} \label{quadratic_case}
\indent 
 
As a first example we consider form factors motivated by an approach of reconstruction of the effective action for quantum gravity based in data obtained via Causal Dynamical Triangulation (CDT). In Ref. \cite{Knorr}, the authors put forward a reverse engineered procedure to reconstruct the effective action starting from an Euclidean template of the form

\begin{align}\label{eff_Action_Knorr}
\Gamma = \frac{2}{\kappa^2} \int d^4x\, \sqrt{g} 
\bigg( 2\Lambda - R - \frac{b^2}{6} R \,\Box^{-2} R \bigg) \,,
\end{align}
and adjusting the free parameter $b$ by matching the autocorrelation of the 3-volume operator with data from CDT.
The same class of effective action has been motivated by cosmological considerations. In fact, in Ref. \cite{Maggiore_1} the authors proposed an effective model with non-localities of the type $R \, \Box^{-2} R$ as an alternative model for dark energy. It can be found an extended version involving non-local term of the type $C_{\mu\nu\alpha\beta}  \, \Box^{-2} C^{\mu\nu\alpha\beta}$ in Ref. \cite{Maggiore_2}. For an up-to-date overview on the various aspects of cosmological evolution driven by this class of non-localities see Ref. \cite{Maggiore_3}. Furthermore, contributions like $R \, \Box^{-2} R$ and $C_{\mu\nu\alpha\beta}  \, \Box^{-2} C^{\mu\nu\alpha\beta}$ were earlier obtained as a consequence of a decoupling mechanism in a renormalization group analysis \cite{Shapiro_Ex1}.

 In the present paper we consider the same type of non-locality appearing in Eq. \eqref{eff_Action_Knorr}, but also including the term $C_{\mu\nu\alpha\beta}  \, \Box^{-2} C^{\mu\nu\alpha\beta}$. In this sense, we consider the following class of form factors
\begin{eqnarray}\label{Form_factor_Ex1}
F(\Box) = - \frac{\rho_0}{\Box^2} \, , 
\qquad \textmd{and} \qquad
W(\Box) = - \frac{\rho_2}{\Box^2} \, , 
\end{eqnarray}
with $\rho_0$ and  $\rho_2 $ being positive parameters. Before proceed, we must clarify some points regarding these form factors. First of all, we note that the reconstruction approach proposed in \cite{Knorr}, in the context of this paper, is simply used as a motivation for choosing the functional form of $F(\Box)$ and $W(\Box)$. Then, we do not impose any restriction on the parameters $\rho_0$ and $\rho_2$ coming from the matching template approach discussed in Ref. \cite{Knorr}. It all important to point out that the effective action in Eq. \eqref{eff_Action_Knorr} is written according to Euclidean signature and the passage to the Lorentizian signature is done by means of ``naive'' Wick rotation. We emphasize, however, that a completely well defined Wick rotation in quantum gravity remains as an open problem and is not addressed here. 

Considering the class of form factors introduced above, as well as the definition of the $Q$-factors defined in Eqs. \eqref{def_Q2} and \eqref{def_Q0}, the relevant integrals contributing to the NR gravitational potential are given by
\begin{subequations}
	\begin{align}
	I^{(s)}_1(r) = \int \frac{d^3\vec{q}}{(2 \pi)^3} \, \frac{1}{\vec{q}^{\,2} + \mu_s^2}\, e^{i \vec{q} \cdot \vec{r}}
	= \frac{e^{-\mu_s r}}{4 \pi r} \label{Knorr_I_1} ,
	\end{align}
	\begin{align}
	I^{(s)}_0(r) = \int \frac{d^3\vec{q}}{(2 \pi)^3} \, \frac{ \vec{q}^{\,2}  }{\vec{q}^{\,2} + \mu_s^2}\, e^{i \vec{q} \cdot \vec{r}} 
	= \delta^3 (\vec{r})- \mu_s^2 \, \frac{e^{-\mu_s r}}{4 \pi r} \label{Knorr_I_0} ,
	\end{align}
	\begin{align}
	I_{ij}^{(s)}(r) &= \int \frac{d^3\vec{q}}{(2 \pi)^3} \, \frac{ \vec{q}_i \vec{q}_j}{  \vec{q}^{\,2} + \mu_s^2 } \,e^{i \vec{q} \cdot \vec{r}} \nonumber \\
	&= \frac{1}{3} \delta_{ij} \delta^3(\vec{r}) 
	+ \bigg\{ (1 + \mu_s r) \delta_{ij}  
	-  (3 + 3 \mu_s r + \mu_s^2 r^2) \frac{x_i x_j}{r^2} \bigg\} \frac{e^{-\mu_s r}}{4 \pi r^3} \, ,
	\end{align}
\end{subequations}
where we define $\mu_s^2 = 2(\rho_s - \Lambda)$. We shall consider $\rho_s > \Lambda$ such that the non-local form factors  \eqref{Form_factor_Ex1} introduce mass terms in the graviton propagator. The resulting contributions to the NR potential are written as follows (throwing away Dirac delta terms)
\begin{subequations}
	\begin{align}
	V_{\textmd{mon-mon}}^{(s=0)}(r) =  
	&-\frac{\kappa^2 m_1 m_2}{24 \pi \,r} \left( e^{-\mu_2 r} - \frac{1}{4}e^{-\mu_0 r}\right) \nonumber \\
	&+\frac{\kappa^2 m_1 m_2}{192\pi \,r} \bigg(\frac{1}{m_1^2} + \frac{1}{m_2^2}\bigg) \left( \mu_2^2 \, e^{-\mu_2 r} + \frac{5}{4} \mu_0^2 \,e^{-\mu_0 r} \right) \,,
	\end{align}
	\begin{align}
	V_{\textrm{vel-vel}}^{(s=0)}(r) =  -\frac{\kappa^2 m_1 m_2\,\vec{p}^{\,2}}{24\pi r} 
	\bigg\{ \bigg(\frac{1}{m_1^2} + \frac{1}{m_2^2}\bigg)  \left( e^{-\mu_2 r} + \frac{1}{8} e^{-\mu_0 r}\right)
	+ \frac{3}{m_1 m_2} e^{-\mu_2 r}  \bigg\} \,,
	\end{align}
\end{subequations}
for spin-0 particles, and
\begin{subequations}
	\begin{align}
	V_{\textmd{mon-mon}}^{(s=1/2)}(r) =  
	-\frac{\kappa^2 m_1 m_2}{24 \pi \,r} \left( e^{-\mu_2 r} - \frac{1}{4}e^{-\mu_0 r}\right) \,,
	\end{align}
	\begin{align}
	V_{\textrm{vel-vel}}^{(s=1/2)}(r) =  -\frac{\kappa^2 m_1 m_2\,\vec{p}^{\,2}}{24\pi r} 
	\bigg\{ \bigg(\frac{1}{m_1^2} + \frac{1}{m_2^2}\bigg)  \left( e^{-\mu_2 r} + \frac{1}{8} e^{-\mu_0 r}\right)
	+ \frac{3}{m_1 m_2} e^{-\mu_2 r}  \bigg\} \,,
	\end{align}
	\begin{align}
	V_{\textrm{spin-orbit}}^{(s=1/2)}(r) &= \frac{\kappa^2 m_1 m_2}{24\pi r^3} 
	\bigg(\frac{1}{m_1^2} \vec{S}_1 \cdot\vec{L} + \frac{1}{m_2^2} \vec{S}_2\cdot\vec{L} 
	+ \frac{3(\vec{S}_1+\vec{S}_2)\cdot\vec{L}}{2\,m_1 m_2} \bigg) 
	 \,  (1+r\mu_2)e^{-\mu_2 r}    \nonumber\\
	&\,+\frac{\kappa^2 m_1 m_2}{192\pi r^3} 
	\bigg(\frac{1}{m_1^2} \vec{S}_1 \cdot\vec{L} + \frac{1}{m_2^2} \vec{S}_2\cdot\vec{L}\bigg) 
	\, (1+r\mu_0)e^{-\mu_0 r}  \, ,
	\end{align}
	\begin{align}
	V_{\textrm{spin-spin}}^{(s=1/2)}(r) = &-\frac{\kappa^2}{32\pi \,r^3} \vec{S}_1 \cdot \vec{S}_2\, (1+r\mu_2+r^2 \mu_2^2) e^{-\mu_2 r} \nonumber \\ 
	&+\frac{\kappa^2}{32\pi \,r^3} (\hat{r}\cdot\vec{S}_1 )\, (\hat{r}\cdot \vec{S}_2)\, (3+3r\mu_2+r^2 \mu_2^2) e^{-\mu_2 r} \,,
	\end{align}
\end{subequations}
in the case of spin-1/2 scattered particles.

As we observe, both monopole-monopole and velocity-velocity sectors are composed exclusively by terms scaling with usual $r^{-1}$ behavior, but with an additional exponential damping as a result of mass-like terms in the graviton propagator. By a simple comparison of $V_{\textmd{mon-mon}}^{(s)}(r)$ and $V_{\textmd{vel-vel}}^{(s)}(r)$ we quickly infer the suppression of velocity-velocity contribution due to the ``overall'' ratio $\vec{p}^{\,2}/(m_im_j)$ ($\ll 1$ in the NR limit). Since both sectors exhibit similar $r$-dependencies, the dominance of $V_{\textmd{mon-mon}}^{(s)}(r)$ over $V_{\textmd{vel-vel}}^{(s)}(r)$ is valid for all distance scales (at least, within our approximations).

Before we move on to spin-dependent contributions, let us have a closer look at the monopole-monopole sector associated with spin-0 particles. As anticipated in the previous section, $V_{\textmd{mon-mon}}^{(s=0)}(r)$ shows an additional contribution beyond the usual terms appearing in the static limit. In the present example, this extra contribution is given by
\begin{align}
\Delta V_{\textmd{mon-mon}}^{(s=0)}(r) =  
\frac{\kappa^2 m_1 m_2}{192\pi \,r} \bigg(\frac{1}{m_1^2} + \frac{1}{m_2^2}\bigg) \left( \mu_2^2 \, e^{-\mu_2 r} + \frac{5}{4} \mu_0^2 \,e^{-\mu_0 r} \right) \,.
\end{align} 
The suppression mechanism regarding this term is readily understood in terms of some physical considerations involving the static limit, namely
\begin{align}
V_{\textmd{static}}(r) =  
-\frac{\kappa^2 m_1 m_2}{24 \pi \,r} \left( e^{-\mu_2 r} - \frac{1}{4}e^{-\mu_0 r}\right) .
\end{align}
In order to avoid significant deviations from the usual Newtonian potential within regions where the later has been experimentally verified, we impose upper bounds on the mass parameters $\mu_2$ and $\mu_0$. A rough estimate is obtained by assuming the Newtonian potential as a faithful description up the solar system radius. Taking solar system radius as $r_\textmd{S} \sim 10 \,\textmd{AU}$, we recover the appropriated Newtonian potential (for $r<r_\textmd{S}$) provided that $\mu_i \,r_\textmd{S} \ll 1$, leading to the rough limit $\mu_{i} \ll 10^{-25} \,\textmd{MeV} $. In this case, the suppression of the extra term $\Delta V_{\textmd{mon-mon}}^{(s=0)}(r)$ occurs as a consequence of the ratios $\mu_i^2/m_j^2$ that are much smaller than one, even if we consider the elementary particles scattering (with masses of order $\sim \textmd{MeV}$).

Concerning the spin-dependent contributions,  $V_{\textrm{spin-orbit}}^{(s=1/2)}(r)$ and $V_{\textrm{spin-spin}}^{(s=1/2)}(r)$, we observe the appearance of terms with different scaling behaviors in comparison with the previously discussed sectors. In particular, we note that spin-orbit sector involves interactions proportional to  $r^{-2}$ and $r^{-1}$ (recall that $\vec{L} \sim \vec{r}$), while spin-spin interactions also involve terms scaling with $r^{-3}$. In all cases we observe the exponential damping as well.

The long-range potential is dominated by $r^{-1}$-terms, which receives contributions from all the sectors investigated in the paper. Nevertheless, even in the set of interactions scaling with $r^{-1}$, the leading order long-ranging contribution corresponds to the usual static term in the monopole-monopole sector. In this case, the remaining $r^{-1}$-terms are suppressed by factors involving $\vec{p}^{\,2}/(m_im_j)$ and $\mu_i^2/m_j^2$.

The situation turns out to be more intriguing in the short-distance regime, since in this case we observe different dominant sectors for spin-0 and spin-1/2 particles. In the spin-0 case, the leading order short-range contribution come from the usual static terms in the monopole-monopole sector, 
\begin{align}
V_{\textmd{short-range}}^{(s=0)}(r) =  
-\frac{\kappa^2 m_1 m_2}{32 \pi \,r} + \cdots
\end{align}
In the case of spin-1/2 particles, on the other hand, the dominant contribution appears with spin-spin interactions, namely
\begin{align}
V_{\textmd{short-range}}^{(s=1/2)}(r) = -\frac{\kappa^2}{32\pi \,r^3} \left( \vec{S}_1 \cdot \vec{S}_2 -3 (\hat{r}\cdot\vec{S}_1 )\, (\hat{r}\cdot \vec{S}_2) \right)  + \cdots\,.
\end{align}
In both cases, the leading order short-range contribution does not involve any parameter associated with the form factors considered in this example, see Eq. \eqref{Form_factor_Ex1}. This fact may be interpreted as direct consequence of the infrared nature of this form factors class.

\subsection{Form factors motivated by FRG approach for quantum gravity} \label{linear_case}
\indent 
In the second explicit example we consider form factors motivated by a recent strategy employed in the functional renormalization group (FRG) approach for asymptotically safe quantum gravity \cite{Bosma_PRL_123}. The main idea is to adopt an expansion of a coarse-grained version of the effective action, $\Gamma_k$, in terms of $k$-dependent form factors, where $k$ stands for an infrared cutoff scale introduced in the realm of the FRG framework. Within this formulation, it is possible to use the FRG-equation in order to derive (integro-differential) flow equations for the form factors \cite{Bosma_PRL_123,Knorr_Form_Factors}. This strategy was applied in the search for an asymptotically safe solution in terms of form factors. After some approximations, the authors of Ref. \cite{Bosma_PRL_123} found a fixed point solution that could be fitted into a simple functional dependence of the form factor $W(\Box)$, namely
\begin{align}
W (\Box) = \frac{\rho}{ \Box + \beta} + w \, ,
\label{W_factor_Bosma} 
\end{align}
with the parameters $\rho$, $\beta$ and $w$ being adjusted according with numerical solutions of the fixed point equations. It is worth to mention that due to approximations employed in Ref. \cite{Bosma_PRL_123}, the form factor associated with the sector $R F(\Box) R$ decouples from the flow equation and it is set to zero at the level of the flowing effective action $\Gamma_k$. 

Keeping this in mind, in this section we mainly focus on the contribution of $W(\Box)$ to the NR potentials. For the sake of simplicity, in this example we set the cosmological constant to zero ($\Lambda = 0$). Furthermore, we should also emphasize that our analysis involve two important assumptions: (i) while the result obtained in Ref. \cite{Bosma_PRL_123} is based on non-perturbative euclidean approach, we consider a naive continuation to Minkowski space-time; (ii) we assume that shape of the form factor $W(\Box)$ remains the same once we integrate down to $k=0$. For these reasons, we explore other regions of the parameters space $\rho$, $\beta$ and $w$ instead of restricting ourselves to the particular values obtained in Ref. \cite{Bosma_PRL_123}.

Taking into account this class of form factors, the relevant integrals contributing to the spin-2 sector of the NR potential involve the following term
\begin{align}
\frac{1}{Q_2} = - \frac{1}{2w} \frac{\vec{q}^{\,2} + \beta}{\left( \vec{q}^{\, 2} + A_+ \right) \left( \vec{q}^{\, 2} + A_- \right)} \, , 
\label{inverse_Q2_Bosma} 
\end{align}
which is possible be mapped, by means of partial fraction decomposition, in the standard integrals reported in the Appendix \ref{Appendix_int}. Note that we define
\begin{align}
A_{\pm} = \frac{\left( - 1 + 2 \rho + 2 \,w \beta \right) 
	\pm \sqrt{ \left( - 1 + 2 \rho + 2 w \beta \right)^2 + 8 w \beta } }{4 w} \, .
\label{def_A_pm} 
\end{align}

\begin{figure}[ht]
	{\includegraphics[width = 2.5in]{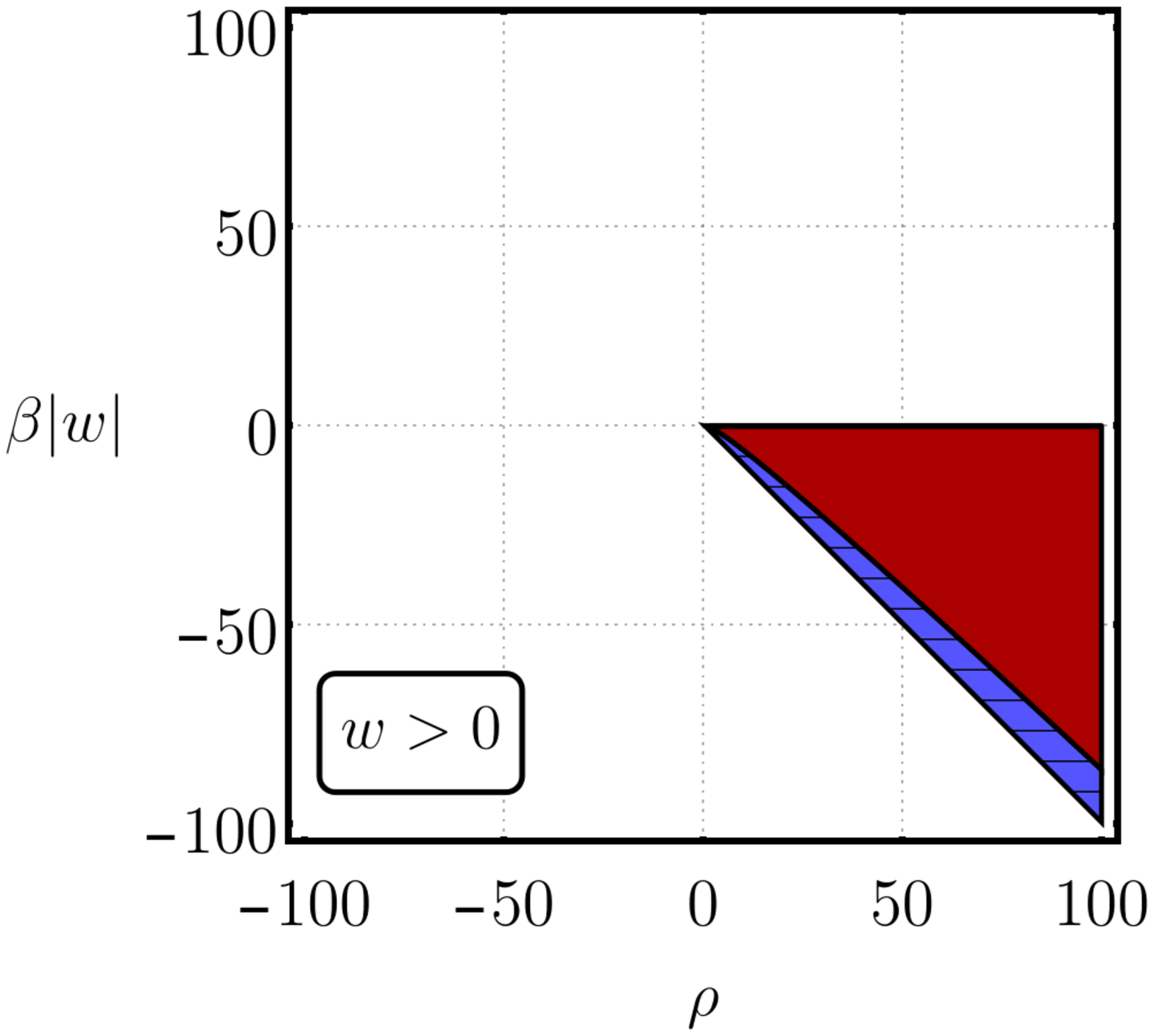}} \qquad
	{\includegraphics[width = 2.5in]{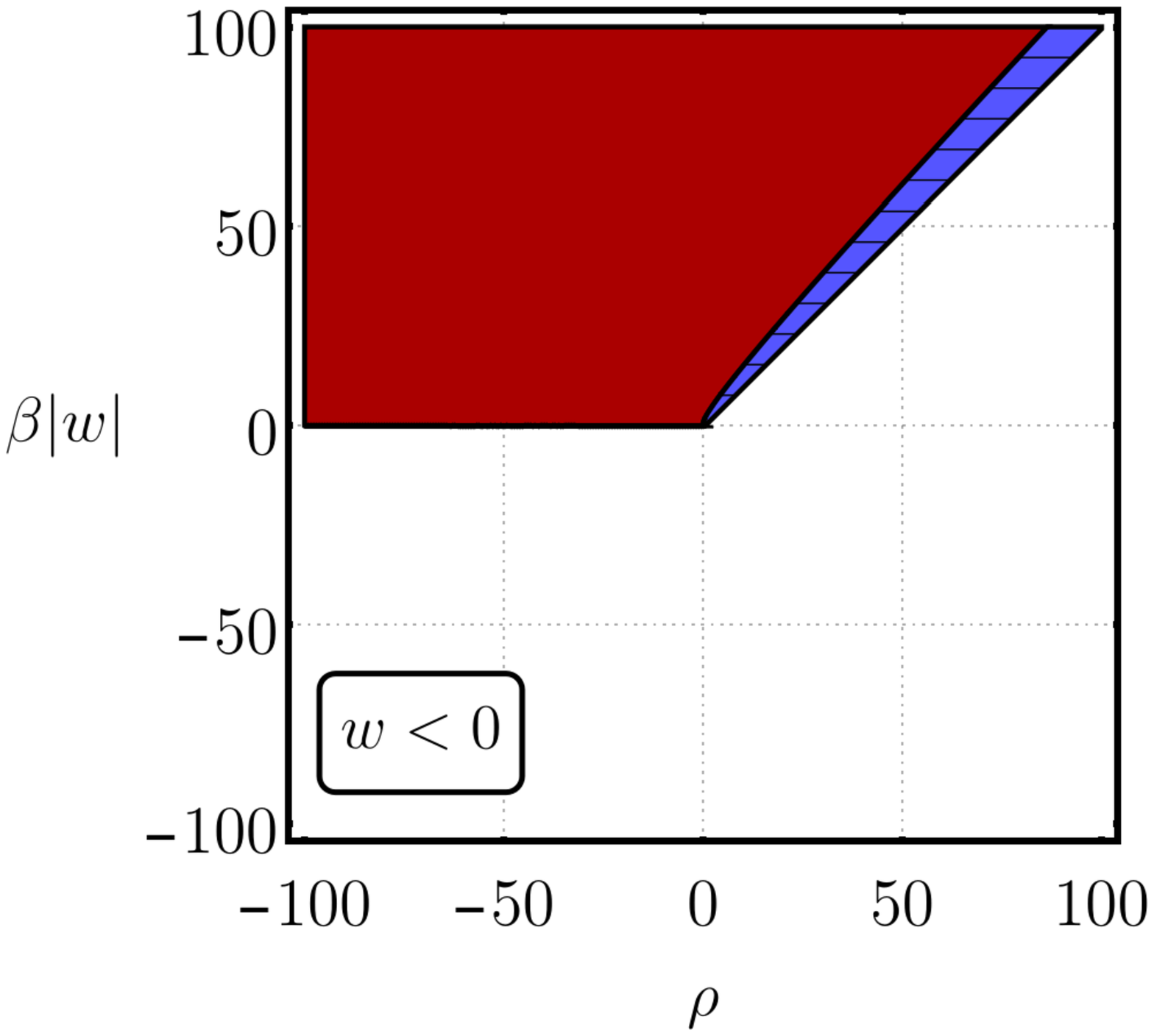}}
	\caption{Regions in the space of parameters ($\rho$, $\beta |w|$) for positive and negative values of $w$. The red regions correspond to the values in which $A_{\pm} \in \mathbb{R}$ and positive (type i). The blue region (with horizontal dashing) indicates values where type-ii restriction is verified ($A_{\pm} \in \mathbb{C} $, such that $\textmd{Re}(A_\pm) > 0$ and $A_\pm^* = A_\mp$).}
	\label{RegionPlots}
\end{figure}

Before we discuss the main results of this section, it is important to observe that an appropriate mapping in terms of the standard integrals \eqref{int_1}-\eqref{int_3}  requires some restrictions on $A_\pm$. Therefore, one must have a closer look at the dependence of $A_{\pm}$ with respect to the parameters $\rho$, $w$ and $\beta$. In particular, we want to probe the existence of regions in the parameters space $\rho$, $w$ and $\beta$ where one of the following conditions is verified
\begin{itemize}
	\item[(i)] $A_{\pm} \in \mathbb{R} $, with $A_\pm > 0$,
	\item[(ii)] $A_{\pm} \in \mathbb{C} $, such that $\textmd{Re}(A_\pm) > 0$ and $A_\pm^* = A_\mp$.
\end{itemize}
In the first case, the resulting  potential is composed by a sum of terms with $r$-dependency characterized by $1/ r^\alpha$ and $e^{- \sqrt{A_\pm}\,r}/ r^\alpha$ (with $\alpha=1,2,3$). When $A_{\pm}$ takes complex values we also observe oscillatory terms (modulated by an exponential dumping) coming from the imaginary part of $A_{\pm}$. In this case, the additional restriction $A_\pm^* = A_\mp$ appears as a reality condition for the resulting potential. In Fig. \ref{RegionPlots} we show the existence of regions in the parameters space defined by $\rho$, $w$ and $\beta$ where the aforementioned conditions are verified. We note that, since the non-trivial dependence of $A_\pm$ occurs with respect to $\rho$ and the quantity $\beta w$, we summarized the results in terms of two region-plots in the plane $(\rho,\beta \,|w|)$. Apart from $\rho$ and $\beta \,|w|$, the shape of viable regions depends on the sign of $w$. Both signs of $w$ admit dense regions satisfying conditions of type-i (red) or type-ii (blue). 

Since the complete expressions are quite long, here we shall not report the full results for $V^{(s=0)}(r)$ and $V^{(s=1/2)}(r)$. Nonetheless, we note that it is directly obtained in terms of Eqs. \eqref{spin_0_pot} and \eqref{spin_meio_pot} by putting together the explicit integrals
\begin{subequations}
	\begin{align}
	I^{(2)}_1(r) =& - \frac{1}{2w} \, \frac{1}{\left( m_+^2 - m_-^2 \right)} \Bigg\{ 
	\frac{ \left( m_+^2 - m_-^2 \right)}{m_+^2 m_-^2} \frac{\beta}{4 \pi r} \nonumber \\ 
	&- \left( 1 - \frac{\beta}{m_+^2} \right) \frac{e^{- m_+ r}}{4 \pi r} + \left( 1 - \frac{\beta}{m_-^2} \right) \frac{e^{- m_- r}}{4 \pi r} \Bigg\} 
	\, , \label{I2_1_Bosma} 
	\end{align}
	\begin{align}
	I^{(2)}_0(r)  =  -\frac{1}{2w} \, \frac{1}{\left( m_+^2 - m_-^2 \right)} \Bigg\{ \left( m_+^2 - \beta \right) \frac{e^{- m_+ r}}{4 \pi r} - \left( m_-^2 - \beta \right) \frac{e^{- m_- r}}{4 \pi r} \Bigg\}
	\, , \label{I2_0_Bosma} 
	\end{align}
	\begin{align}
	I^{(2)}_{ij}(r) = &- \frac{1}{2w} \, \frac{1}{\left( m_+^2 - m_-^2 \right)} \Bigg\{ 
	\beta \, \frac{ \left( m_+^2 - m_-^2 \right)}{m_+^2 m_-^2}  \left( \delta_{ij} - 3 \frac{x_i x_j}{r^2} \right) \frac{1}{4 \pi r^3} 
	 \nonumber \\
	&- \left( 1 - \frac{\beta}{m_+^2} \right) \bigg[  \left( 1 + m_+ r\right) \delta_{ij} -\left( 3 + 3 m_+ r + m_+^2 r^2 \right) \frac{x_i x_j}{r^2} \bigg] \frac{e^{- m_+ r}}{4 \pi r^3}  \nonumber \\
	&+ \left( 1 - \frac{\beta}{m_-^2} \right) \bigg[ \left( 1 + m_- r\right) \delta_{ij} - \left( 3 + 3 m_- r + m_-^2 r^2 \right) \frac{x_i x_j}{r^2}  \bigg] \frac{e^{- m_- r}}{4 \pi r^3}  \Bigg\}
	\, ,\label{I2_ij_Bosma} 
	\end{align}
\end{subequations}
in which $m_\pm = \sqrt{A_\pm}$ and we assume $m_+^2-m_-^2 \neq 0$. In this case, the contact terms ($\sim \delta^3(\vec{r})$) resulting from integrals of the form \eqref{int_2} and \eqref{int_3} completely cancel out in the final expression. The scaling of the different sectors contributing to the NR potentials is summarized in Table \ref{Table_1}.

\begin{table}[h]
	\begin{tabular}{|c|c|c|c|c|c|c|}
		\hline
		\hline
		& $\quad\,\, 1/r \quad\,\,$     & $\quad\, 1/r^{2} \quad\,$ & $\quad\, 1/r^{3} \quad\,$     & $\,\,e^{-m_\pm r}/r\,\,$ & $\,\,e^{-m_\pm r}/r^2\,\,$ & $\,\,e^{-m_\pm r}/r^3\,\,$ \\\hline
		mon-mon                                        & $\checkmark_{0,\,1/2}$ &          &              & $\checkmark_{0,\,1/2}$       &                    &                    \\\hline
		vel-vel                                        & $\checkmark_{0,\,1/2}$ &          &              & $\checkmark_{0,\,1/2}$       &                    &                    \\\hline
		$\vec{L}\cdot\vec{S}_{1,2}$  &                    &          & $\checkmark_{1/2}$ &                    & $\checkmark_{1/2}$       & $\checkmark_{1/2}$       \\\hline
		$\vec{S}_1\cdot\vec{S}_2$                        &              &          & $\checkmark_{1/2}$ & $\checkmark_{1/2}$       & $\checkmark_{1/2}$       & $\checkmark_{1/2}$       \\\hline
		$(\hat{r}\cdot\vec{S}_1)(\hat{r}\cdot\vec{S}_2)$ &           &        & $\checkmark_{1/2}$ &           $\checkmark_{1/2}$   &  $\checkmark_{1/2}$ & $\checkmark_{1/2}$  \\  
		\hline
		\hline 
	\end{tabular}
    \caption{Scaling behavior of the different sectors contributing to the NR potentials $V^{(s=0)}(r)$ and $V^{(s=1/2)}(r)$. The subscript indicates if the correspondent behavior appears for spin-0 and/or spin-1/2 cases.}
    \label{Table_1}
\end{table}

In the static regime the only remaining contribution comes from the monopole-monopole sector, resulting in the following expression
\begin{align}\label{Static_Pot_Ex2}
V_{2,\textmd{static}}(r) = -\frac{\kappa^2 m_1 m_2}{24\pi r} 
\left( 1 - \frac{1}{2w} \frac{1-\beta/m^2_{-}}{m^2_{+}-m^2_{-}} e^{-m_{-}r} +
\frac{1}{2w} \frac{1-\beta/m^2_{+}}{m^2_{+}-m^2_{-}} e^{-m_{+}r}  \right) \,,
\end{align}
with the subscript ``2'' indicating we count only spin-2 contributions in the graviton propagator. Deviations from the $1/r$-behavior within experimentally tested scales is avoided when $m_{\pm} r_{\textmd{min}} \gg 1$, with $r_{\textmd{min}}$ being the smaller distance in which the Newtonian $1/r$-law is validated (see Refs. \cite{Short_distance_1,Short_distance_2,Short_distance_3} for short-distance probes of the Newtonian potential). In this case, the exponential factors strongly suppress the second and third term in Eq. \eqref{Static_Pot_Ex2} and the large distance behavior is dominated by the $1/r$-contribution. As it is noted in Ref. \cite{Bosma_PRL_123}, the static potential in Eq. \eqref{Static_Pot_Ex2} has a particularly interesting behavior at small distances. In this regime, the $1/r$ terms cancel out among different contributions, resulting in a finite potential at $r=0$. 

Beyond the static limit, the NR potential receives multiple contributions scaling with different $r$-dependencies as it is summarized in the Table \ref{Table_1}. In the large distance regime, even if we include contribution beyond the monopole-monopole sector, the leading order term, in the spin-0 and spin-1/2 cases, corresponds to the usual $1/r$ decay, 
\begin{align}\label{Long_range_Ex2}
V_{2,\,\textmd{long-range}}(r) = -\frac{\kappa^2 m_1 m_2}{24\pi r} + \cdots \,.
\end{align}
In this limit, all remaining terms are suppressed either by exponential decay (with $m_{\pm} r \gg 1$) or by sub-leading behavior of $1/r^3$ in comparison with $1/r$. 

In the short-distance regime we observe more intriguing features once contributions beyond of monopole-monopole sector are taken. In this case, the leading order terms are given by
\begin{subequations}
	\begin{align}\label{Short-range_Ex2_s=0}
	V_{\textmd{short-range}}^{(s=0)}(r) =  
	\frac{\kappa^2\, m_1 m_2}{384 \pi w \,r} \left( \frac{1}{m_1^2} + \frac{1}{m_2^2}\right) + \cdots \,,
	\end{align}
	\begin{align}\label{Short-range_Ex2_s=1/2}
	V_{\textmd{short-range}}^{(s=1/2)}(r) =  
    -\frac{\kappa^2}{128\pi w\,r} \left( \vec{S}_1 \cdot \vec{S}_2 +  
    (\hat{r}\cdot\vec{S}_1)(\hat{r}\cdot\vec{S}_2) \right) + \cdots .
	\end{align}
\end{subequations}
Similarly to the example explored in the previous section, we also note different leading order contributions to $V_{\textmd{short-range}}^{(s=0)}(r)$ and $V_{\textmd{short-range}}^{(s=1/2)}(r)$. However, a different aspect in the present case is that the leading order terms at small distances exhibit a dependence with respect to the form factor $W(\Box)$ due to the parameter $w$ in Eqs. \eqref{Short-range_Ex2_s=0} and \eqref{Short-range_Ex2_s=1/2}. This fact indicates that the form factor studied along this section plays an important role the in the UV aspect of the NR potential, even in the presence of terms beyond the monopole-monopole sector. Despite of this fact, our results, for short-range regime,  point out an important difference with respect to the static case, Eq. \eqref{Static_Pot_Ex2}, namely, the cancellation of the Newtonian singularity at $r=0$ does not survive beyond the static limit.


\section{Remarks on the cancellation of Newtonian singularities} \label{Singularities}

The observation at the end of the previous section trigger a question regarding the cancellation of Newtonian singularities. In particular, it would be interesting to investigate whether the regular behavior at $r=0$, observed in higher-derivative models of gravity \cite{Stelle,Accioly_et_al_PRD2018,Cancellation_1,Cancellation_2}, persists after the inclusion of contributions beyond the static limit. This particular test is easily addressed in terms of the results presented in Section \ref{Sec_Potentials}, however, it requires a slight modification in the way we interpret our framework. In the case of higher-derivative models, the form factor expansion appears at the level of the classical action \cite{Accioly_et_al_PRD2018,Cancellation_1,Cancellation_2}, given by
\begin{align}
S_{\textmd{HD}}[g_{\mu \nu}] = 
\frac{2}{\kappa^2} \int d^4 x \, \sqrt{-g} \left(  -  R - \frac{1}{3} R F(\Box) R + C_{\mu \nu \alpha \beta} W(\Box) C^{\mu \nu \alpha \beta} \right) + \mathcal{O}(\mathcal{R}^3) \,,
\label{Class_action_HD} 
\end{align}
with polynomial form factors ($p,q \in \mathbb{N}$)
\begin{align}\label{Form_Factor_Poly}
F(\Box) = \sum_{n=0}^p f_n \,(-\Box)^n \qquad \textmd{and} \qquad W(\Box) = \sum_{n=0}^q w_n \,(-\Box)^n \,.
\end{align}
In this case, all the results presented in Section \ref{Sec_Potentials} remain unchanged, however, keeping in mind that Eq. \eqref{prop_simpl} should be interpreted as the tree-level graviton propagator.

The relevant integrals appearing in Eqs. \eqref{spin_0_pot} and \eqref{spin_meio_pot} are computed by means of the partial decomposition
\begin{align}\label{Decomposition_HD}
\frac{1}{\vec{q}^{\,2}\,Q_a(-\vec{q}^{\,2})} = \frac{1}{\vec{q}^{\,2}} + 
\sum_{i=1}^{\mathcal{N}_a} \frac{\mathcal{R}_{i}^{(a)}}{\vec{q}^{\,2}+\mu^2_{a,i}}  \, , \qquad
\textmd{with $a=0,2$}\,,
\end{align}
where $\mathcal{N}_a = p\,\delta_{a,0}+q\,\delta_{a,2}+1$ and we define the residues
\begin{align}
\mathcal{R}_{n}^{(a)} = 
-\prod_{ \substack{l = 1\\ l \neq n} }^{\mathcal{N}_a}
\frac{ \mu^2_{a,l} }{ \mu^2_{a,l} - \mu^2_{a,n} } \,.
\end{align}
The mass parameters $\mu_{a,l}$ are defined as the zeros of the $Q_a$-factors, namely $Q_a(\mu^2_{a,l})=0$. In order to avoid complications with degenerate poles we assume $\mu^2_{a,i} \neq 0$ and $ \mu^2_{a,i} \neq \mu^2_{a,j}$ if $i\neq j$. In such a case, we can decompose $I_n^{(a)}(r)$ and $I_{ij}^{(a)}(r)$ in terms of the standard integrals \eqref{int_1}-\eqref{int_3}, as displayed below
\begin{subequations}
	\begin{align}\label{Int_HD_1}
	I_n^{(a)}(r) = \mathcal{I}_n(r,0)+
	\sum_{l=1}^{\mathcal{N}_a} \mathcal{R}_{l}^{(a)} \,\mathcal{I}_n(r,\mu_{a,l})\,,
	\end{align}
	\begin{align}\label{Int_HD_2}
	I_{ij}^{(a)}(r) = \mathcal{I}_{ij}(r,0)+
	\sum_{l=1}^{\mathcal{N}_a} \mathcal{R}_{l}^{(a)} \,\mathcal{I}_{ij}(r,\mu_{a,l}) \,.
	\end{align}
\end{subequations}

The explicit NR potentials are obtained by using Eqs. \eqref{Int_HD_1} and \eqref{Int_HD_2}. The scaling dependence of the different sectors exhibits the same behavior of the previous subsection (see Table \ref{Table_1}).

The static limit (Eq. \eqref{pot_static}) has been computed before, e.g. see Refs. \cite{Accioly_et_al_PRD2018,Cancellation_1,Cancellation_2}, resulting in the following expression
\begin{align}
V_{\textmd{static}}(r) &=  
-\frac{\kappa^2 m_1 m_2}{32\pi \,r} \bigg( 
1 + \frac{4}{3}
\sum_{l=1}^{q+1} \mathcal{R}_{l}^{(2)} \,e^{-\mu_{2,l} r}  - \frac{1}{3}
\sum_{l=1}^{p+1} \mathcal{R}_{l}^{(0)} \,e^{-\mu_{0,l} r} \bigg)  \,, \nonumber\\
&\!\! \underset{r\to 0}{=} -\frac{\kappa^2 m_1 m_2}{32\pi \,r} \bigg( 
1 + \frac{4}{3}
\sum_{l=1}^{q+1} \mathcal{R}_{l}^{(2)}  - \frac{1}{3}
\sum_{l=1}^{p+1} \mathcal{R}_{l}^{(0)}  \bigg) + \textmd{finite} \,.
\end{align}
The cancellation of the $1/r$ singularity follows from the property $\sum_{l=1}^{\mathcal{N}_a} \mathcal{R}_{l}^{(a)} = -1$ (see Ref. \cite{Cancellation_2}).

Taking contributions beyond the static sector, the regularity of the NR potential at $r=0$ becomes more subtle. As an example, we consider the particular case corresponding to Stelle's Quadratic Gravity ($p=q=0$) \cite{Stelle}. In such a case, the leading order short-distance contribution is given by
\begin{subequations}
	\begin{align}\label{Short-range_Stelle_s=0}
	V_{\textmd{Stelle}}^{(s=0)}(r) =  -
	\frac{\kappa^2\, m_1 m_2 }{192 \pi  \,r} \left( \frac{1}{m_1^2} + \frac{1}{m_2^2}\right) 
	\left( \mu_{2}^2 + \frac{5}{4} \mu_{0}^2 \right) + \textmd{finite} \,,
	\end{align}
	\begin{align}\label{Short-range_Stelle_s=1/2}
	V_{\textmd{Stelle}}^{(s=1/2)}(r) =  
	\frac{\kappa^2\,\mu_{2}^2}{64 \pi  \,r} \left( \vec{S}_1 \cdot \vec{S}_2 +  
	(\hat{r}\cdot\vec{S}_1)(\hat{r}\cdot\vec{S}_2) \right) + \textmd{finite}  .
	\end{align}
\end{subequations}
We note quite a similar behavior in comparison with the example of the previous section. As we can observe, the $1/r$ singularity reappears once we include contributions beyond the static limit. This result indicates that additional UV modifications should be included in order to keep the NR potential finite at $r=0$. Indeed, this is actually the case as one can easily see by taking into account higher terms in the polynomial form factor defined in Eq. \eqref{Form_Factor_Poly}. A simple example is the sixth-order higher-derivative gravity ($p=q=1$) which results in a singularity-free potential, even after the inclusion of contributions beyond the static limit. The same behavior is also observed for any $p,q\geq1$. A similar conclusion for the cancellation of singularities at $r=0$, but at the level of the Kretschmann scalar, was obtained in Refs. \cite{Breno_Tiberio_1}. It is important to reinforce that the discussion presented here is restricted to the cancellation of the Newtonian singularity in the classical (tree-level) contribution to the NR potential and, therefore, our results should not be interpreted as a definitive claim concerning the problem of singularity resolution.


\section{Concluding Comments} \label{Sec_Concluding}

In this paper, we investigate quantum effects in the NR gravitational inter-particle potential, including contributions beyond the static regime. We consider both the gravitational scattering of spin-0 and spin-1/2 particles. Our results are based on the form factor expansion of the effective action in the covariant approach for quantum gravity. Within this formalism, the quantum corrections are encoded in the form factors $F(\Box)$ and $W(\Box)$ associated with curvature squared terms in the effective action. Considering metric fluctuations around flat background, these form factors capture all the relevant information concerning the (flat) graviton propagator. Our main results are summarized as follows:
\begin{itemize}
    \item In the monopole-monopole sector, the NR potentials associated with spin-0 and -1/2 particles exhibit a universal leading-order contribution but differ with respect to a sub-leading term. The velocity-velocity sector exhibits the same result for spin-0 and spin-1/2 particles.
    \item The NR potential associated with the scattering of spin-1/2 particles also involves spin-orbit and spin-spin interactions. We observe that the form factors $F(\Box)$ and $W(\Box)$ may contribute to spin-orbit, while only $W(\Box)$ can generate corrections to spin-spin interactions.
    \item Comparing our results with previous investigations concerning the electromagnetic NR potential, we observe similar interaction structures appearing in both cases. 
\end{itemize}

We apply the results obtained in Sec. \ref{Sec_Potentials} to explicit examples of form factors motivated by non-perturbative approaches for quantum gravity. In the first example, we consider form factors motivated by an approach where the effective action  was obtained by matching a predefined template with CDT data \cite{Knorr}. In the second  one, we explore a form factor obtained in the context of the FRG approach for asymptotically safe quantum gravity \cite{Bosma_PRL_123}. The two cases have the contributions to the NR potentials reduced to the form $1/r^{\alpha}$ or $e^{-m r}/r^{\alpha}$ with $\alpha = 1,2,3$. Furthermore, the dominant short-range contributions depend on the type of particle being scattered.

The example studied in Sec. \ref{linear_case} presents the reappearance of the singularity at $r=0$ once contributions beyond the static regime are accounted. Motivated by this result, in Sec. \ref{Singularities} we revisit the cancellation of Newtonian singularities in higher-derivative models. Within  this class of models, our results indicate that the cancellation of singularities at $r=0$ requires a higher number of derivatives when compared with the static approximation.

The  analysis performed here  only includes quantum corrections at the level of the graviton propagator, while the vertices are taken to be tree-level ones. This is an important approximation in our approach and deserves further investigation. In principle, we could also adopt a form factor expansion in order to capture quantum corrections at the level of gravity-matter systems (see, for example, Ref. \cite{Knorr_Form_Factors}). However, this approach increases considerably the  calculations of the inter-particle potentials.  In a recent work \cite{Draper}, a remarkable progress was made by taking into account the most general parameterized relativistic amplitude for gravity-mediated scattering of scalar particles.
 
 In addition, we only consider the scattering of spin-0 and spin-1/2 particles, but we could also include the scattering of spin-1 particle. As discussed in Ref. \cite{HR_0802.0716}, the NR potentials associated with spin-1 scattered particle exhibit new interactions involving the polarization, besides the velocity- and spin-dependent contributions. These points remain to be investigated  in a future work.

\section*{Acknowledgements}
\noindent 
We would like to thank  J.A. Helayël-Neto, J.T. Guaitolini Junior and P.C. Malta for reading the manuscript and the constructive comments. GPB is grateful for the support by CNPq (Grant no.~142049/2016-6) and thanks the DFQ Unesp-Guaratinguet\'a for the hospitality. LPRO is supported by the PCI-DB funds (CNPq/MCTIC). MGC is supported by CNPq funds.

\appendix

\section{Integrals} \label{Appendix_int}
\noindent 

Along this paper we use the following definitions
\begin{equation}
I^{(a)}_n (r) = \int \frac{d^3\vec{q}}{(2 \pi)^3} \, \frac{ e^{i \vec{q} \cdot \vec{r}}   }{  (\vec{q}^{\,2})^n \, Q_a} \, , 
\label{I_a_n} 
\end{equation}
\begin{equation}
I^{(a)}_{ij} (r) = \int \frac{d^3\vec{q}}{(2 \pi)^3} \, \frac{ e^{i \vec{q} \cdot \vec{r}}  }{  \vec{q}^{\,2} \, Q_a} \, \vec{q}_i \vec{q}_j \,,
\label{I_ij} 
\end{equation}
where $n \in \mathbb{N}$ and $ a=0,2$. 


 We highlight that, for functions of type $Q_a = Q_a(\vec{q}^{\,2})$, one can use spherical coordinates and solve the angular part of the Fourier integral \eqref{I_a_n} (see, for instance, Ref. \cite{Accioly_PRD_93}), which leads to a result with only radial dependence. This allows us to recast some integrals. For example,   
\begin{equation}
\int \frac{d^3\vec{q}}{(2 \pi)^3}  \, \frac{ e^{i \vec{q} \cdot \vec{r}}  }{ \vec{q}^{\,2} \, Q_a} \,  i \vec{A} \cdot \vec{q} =
\vec{A} \cdot \vec{\nabla} \left[ I^{(a)}_1 (r) \right] = \vec{A} \cdot \frac{\vec{r}}{r} \, \frac{d}{dr} \, I^{(a)}_1 (r) \, ,
\label{int_A} \end{equation}
with $\vec{A}$ being  a $\vec{q}$-independent vector.  


In special, it is useful to have in mind some particular cases corresponding to standard integrals appearing along the calculations performed in Sec. \ref{Sec_App}, namely
\begin{align} 
\mathcal{I}_1(r,m) \equiv 
\int \frac{d^3\vec{q}}{(2 \pi)^3} \, \frac{ e^{i \vec{q} \cdot \vec{r}}   }{\vec{q}^{\,2} + m^2} = \frac{e^{-mr}}{4 \pi r} \, ,
\label{int_1} 
\end{align}

\begin{align}
\mathcal{I}_0(r,m) \equiv 
\int \frac{d^3\vec{q}}{(2 \pi)^3} \, e^{i \vec{q} \cdot \vec{r}} \, \frac{ \vec{q}^{\,2}   }{\vec{q}^{\,2} + m^2} = 
\delta^3(\vec{r}) - \frac{m^2}{4 \pi r} e^{-mr} \, ,
\label{int_2} 
\end{align}

\begin{align}
\mathcal{I}_{ij}(r,m) &\equiv 
\int \frac{d^3\vec{q}}{(2 \pi)^3} \, e^{i \vec{q} \cdot \vec{r}} \, \frac{ \vec{q}_i  \vec{q}_j }{\vec{q}^{\,2} + m^2}  \nonumber \\
&=\frac{1}{3} \delta_{ij} \delta^3(\vec{r}) + \bigg\{ (1 + mr) \delta_{ij}  
-  (3 + 3mr + m^2 r^2) \frac{x_i x_j}{r^2} \bigg\} \frac{e^{-mr}}{4 \pi r^3} \, .
\label{int_3} 
\end{align}


\end{document}